\documentclass[prb,aps,amsfonts,amssymb,floatfix,showpacs,superscriptaddress]{revtex4}
\usepackage{graphicx}
\usepackage{dcolumn}
\usepackage{bm}
\usepackage{color} 
\usepackage{epsfig}
\usepackage{epstopdf}
\usepackage{amsmath}
\begin{document} 
\title{A model for intertwined orders in cuprates}
\author{R.S. Markiewicz, M. Matzelle, and A. Bansil}
\affiliation{Physics Department, Northeastern University, Boston MA 02115, USA}
\affiliation{Quantum Materials and Sensing Institute, Northeastern University, Burlington, MA 01803, USA}
\begin{abstract}
We model the intertwined orders in the cuprate pseudogap as a textured antiferromagnet (AFM), where the texture arises from confining competing phases on topological defects, i.e., arrays of AFM domain walls.  Three branches of texture are found, which can be interpreted as a strongly frustrated remnant of an underlying eutectoid phase diagram.  This model can describe many key features of intertwined orders in cuprates, including the trisected superconducting dome, and provides clear evidence for a doping/hopping-parameter-dependent Mott-Slater transition in cuprates.

\end{abstract} 
\maketitle

\section{Introduction}
Unraveling the nature of the pseudogap in cuprates has been a major goal in the field for many years, with implications for the mechanism of high-$T_c$ superconductivity.  A related hope is that improved understanding of the cuprate pseudogap can shed light on similar phenomena found in many other correlated materials, including pnictides, nickelates, Moir\'e materials, and organic and heavy fermion compounds\cite{pnpg,nipg,Mopg,orpg,hepg,unpg}.  We have recently demonstrated that several key features of the pseudogap -- including Fermi arcs and a logarithmically-diverging heat capacity at the pseudogap collapse, accompanied by the anomalous growth of the Hall number\cite{Taill1} -- can be understood in terms of the termination of an antiferromagnetic (AFM) phase.\cite{Paper1}  However, this is only half of the story.  A key question is why does the pseudogap not develop long-range order, and, in a similar vein, what is the origin of the intertwined orders\cite{IT,ves} seen in different parts of the pseudogap phase diagram.  Here we assemble various lines of evidence to show that the data can be understood in terms of textures of the AFM phase.

To better understand the picture we have in mind, it is convenient to first recall a more familiar analogue.  Vortices are topological defects of a superconductor, stabilized by a circulating flow of charge, but at the cost of suppressing superconductivity in the vortex core.\cite{Tinkham}  While metastable in a pure superconductor, they play an essential role when a perturbation is applied which competes with superconductivity.  For a type 2 superconductor in a magnetic field, each vortex traps a quantum of magnetic flux, allowing superconductivity to persist to much higher fields than otherwise possible.

Similarly, in undoped cuprates at low temperatures, the AFM order has a small Ising anisotropy, so the topological defects are antiphase boundaries of the AFM order, forming domain walls.\cite{Tranq1}  Metastable at zero doping, they can trap doped charges and preserve AFM order to much higher doping than would otherwise be possible, in the form of stripe phases\cite{Tranq1}.  In one cuprate, La$_{2-x}$ Ba$_x$CuO$_4$, this stripe phase is strong enough to quench superconductivity\cite{Axe}, but in other cuprates, evidence for more fluctuating stripes has been found, but only at low doping, whereas at higher doping they are replaced by charge-density waves (CDWs).  For this reason, stripes are usually considered as secondary manifestations, and not as essential to the pseudogap phase. 

However, we here demonstrate that the stripe-to-CDW transition is a (non-Landau) transition of the dressed defects in this mixed phase, a manifestation of the Mott-Slater transition\cite{MBMB}.  Whereas in the Mott phase the stripes are insensitive to the Fermi surface, in the Slater phase they become dominated by Fermi surface nesting.  Moreover, the CDW nests the AFM Fermi surface, which plays such a large role in Ref.~\onlinecite{Paper1}, and not the nonmagnetic Fermi surface.

The theory of the cuprates is difficult because one must first develop the theory of the homogeneous phase diagram, and then incorporate the effects of the heterogeneous textures.  In Ref.~\onlinecite{Paper1} we set the stage for this analysis, postulating that the pseudogap represents a phase of mostly short-range antiferromagnetic (AFM) order, containing charged domain walls.  In the present paper we extend these results, focusing on the intrinsic heterogeneity of nanoscale phase separation (NPS) and stripe phases.  We show that the three competing stripe textures bear a close resemblance to the eutectoid phase in such materials as Fe-C alloys,\cite{steel} and identify the end phase near the Van Hove singularity (VHS) doping as a nematic phase.

This paper is organized as follows.  In Section II we introduce the three experimentally observed textures of the pseudogap, and demonstrate that the cuprates form an electronic eutectoid system resembling the Fe-C phase diagram.  We show that the nesting vector of the CDW-like stripe branch is controlled by hot-spot nesting of an {\it AFM Fermi surface}, thereby completing our proof that these are textures of an underlying AFM phase.  Supplementary Material (SM) I demonstrates how these three textures describe the main experimental features of the pseudogap.  

Section III shows that the NPS end phase near the VHS is a nematic phase, leading to a close connection between the various electronic textures and the unusual structural physics associated with the low-temperature tetragonal (LTT) phase and suggesting a possible origin for the thermal Hall effect in these materials. In Section IV we show how superconductivity arises in this model.  Section V discusses evidence for a Mott-Slater transition in the pseudogap.  In Section VI we discuss extensions of these results, including consequences for strange metal physics, comparison with density-functional theoy (DFT) results, and a summary of the key features of our model, while Section VII presents our Conclusions.

SM II reviews the many-body underpinning of these calculations, SM III discusses technical details in generating the figures, while SM IV expands on the discussion of various aspects of the pseudogap model.

\section{NPS model of frustrated alloys}
\begin{figure}
\leavevmode
\rotatebox{0}{\scalebox{0.50}{\includegraphics{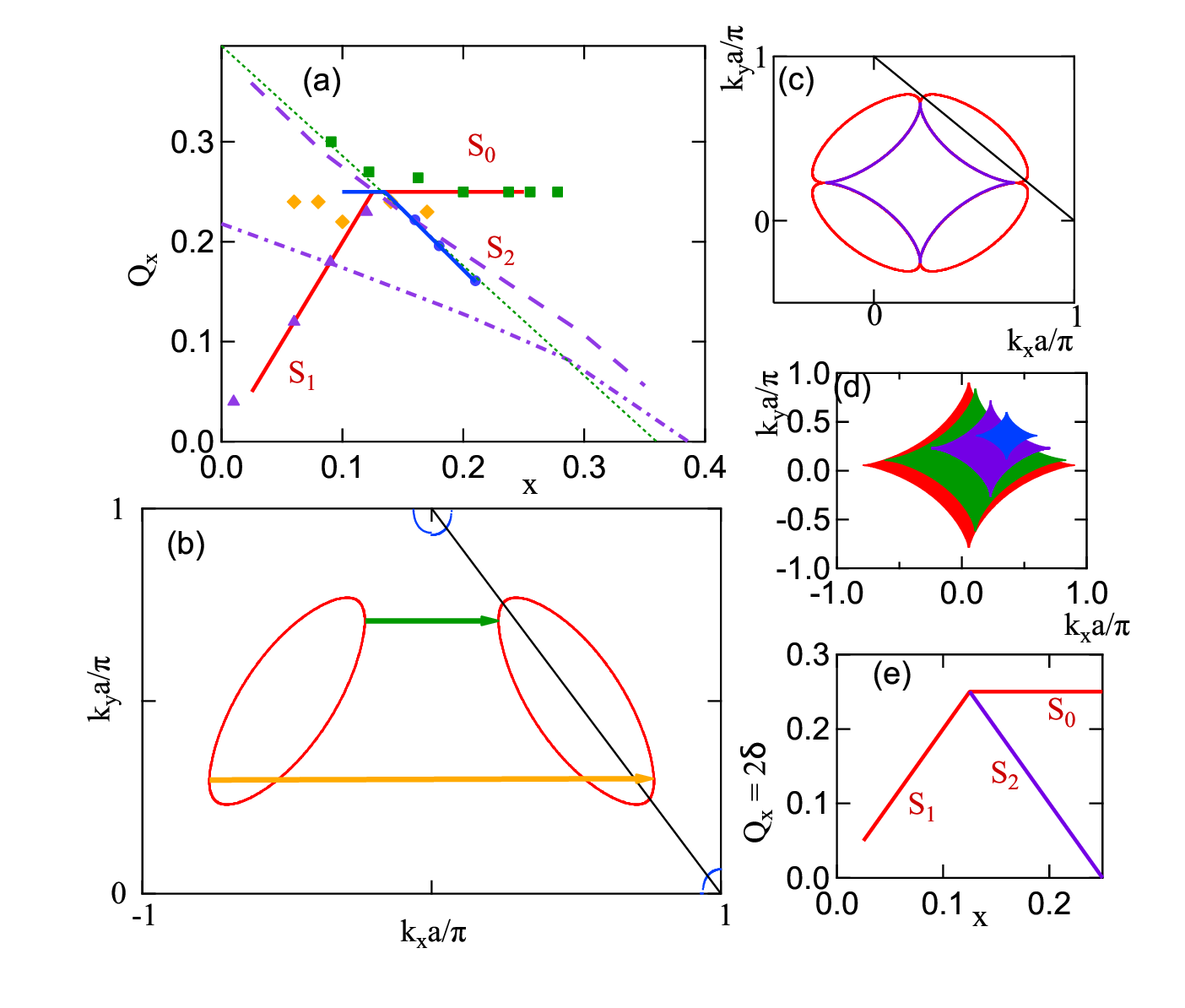}}}
\vskip0.5cm
\caption{(Color online)
(a) Charge order wave vectors $Q_x$, in units of $2\pi/a$, for Bi2201 (purple triangles\cite{TranqBi} and blue circles\cite{EHud2}) and Bi2212 (orange diamonds\cite{per4} and green squares\cite{GhirQ}).  Note that the orange diamonds represent the average of $Q_x$ and $Q_y$, although the differences are small.  Green dotted line is a guide for the eye for the $S_2$ branch.  Thick violet lines represent models of Fermi surface nesting for NM phase (dot-dashed line) or AFM hot-spot nesting (dashed lines). (b-d)  {Hot-spot nesting for AFM with gap $\Delta=$150~meV.}  (b) Nesting vectors $Q$ (green arrow) and $S$ (orange arrow).  (c) Double nesting, leading to star-shaped Fermi surface (violet).  (d)  Evolution of Fermi surface area with doping, $\Delta$ = 250 (blue), 150 (violet), 50 (green), and 20~meV (red).  (e) NPS model of the three textures, reproduced from Ref.~\onlinecite{RSMstr}.
}
\label{fig:6a}
\end{figure}

In discussing the physics of complex systems, such as the cuprate pseudogap, Steve Kivelson and his daughter noted that, ``there is no consensus as to what it would mean to `solve’ the problem, nor what metrics should be used to judge such a solution", and moreover, ``a proper theory cannot possibly account for all the complex phenomena observed in these materials".  Thus, ``a successful theory [must] identify aspects of observable phenomena that it itself identifies as essential."\cite{Kivelson2}  As a first step, we identify the key experimental signatures of intertwined order, Fig.~\ref{fig:6a}(a); understanding these signatures is the main objective of our paper.  Figure~\ref{fig:6a}(a) summarizes the various textures that have been found in the pseudogap phase, identified by their $Q$-vectors.  The data are representative of the Bi-cuprates, specifically Bi$_2$Sr$_2$Ca$_{n-1}$Cu$_n$O$_{6+2n}$ (Bi22(n-1)n), with $n$ = 1,2.  Briefly, branch $S_1$ represents the stripe phases\cite{TranqBi}, $S_0$ is the particular commensurate stripe phase associated with quantum oscillations (QOs) of a Fermi surface much smaller than expected for an AFM \cite{Seba,Ramshaw,per4}, and the $S_2$ branch is associated with the incommensurate CDWs found in both scanning tunneling microscopy (STM)\cite{EHud1,EHud2} and resonant inelastic x-ray scattering (RIXS)\cite{GhirQ} experiments.  In SM I, we show that these branches capture the essential phenomenology of the pseudogap phase.

\subsection{Two routes to strong correlations}
In many-body theory, correlations are defined as any effects beyond Hartree-Fock. Hence, there can be several distinct roads to strong correlations. Most often, strong correlations are discussed in terms of interactions being much stronger than kinetic energy.  This would include the quantum Hall effect of 2D electrons in a magnetic field, heavy Fermion physics, topological surface states, flat bands, strongly renormalized materials with renormalization factors $Z<<1$ including marginal Fermi liquids with $Z\rightarrow 0$ near the Fermi energy, twisted multilayer systems, frustrated electrons in Kagome and other materials, and electrons near a VHS.  Cuprates seem to be an exception to this rule.  Thus, while pnictide superconductors seem in many ways to be conventional metals, only ``moderately correlated" with $Z\sim 1/3$, cuprates are strongly correlated, but with $Z\sim 1/2>> 0.1$ for heavy Fermion compounds. ARPES experiments have searched for signs of marginal Fermi liquid behavior\cite{marginal}, but instead find `low-energy kinks' (LEKs) discussed below.

Instead, one might start from the fact that cuprates are Mott insulators, and as such satisfy the Mott criterion,\cite{MottCrit} that a metal-insulator transition occurs at a critical doping $n_c$ given by
\begin{equation}
a_B(n_c)^{1/3} \sim 0.25,
\label{eq:0}
\end{equation}
where $a_B$ is a material-dependent effective Bohr radius.  This criterion is satisfied by a wide range of materials, including complex materials like cuprates, vanadates, and SrTiO$_3$ as well as doped semiconductors, many of which, including Si and Ge, become superconductors on doping.\cite{MottCrit2}  Is there some property that most of these materials share that can give us a clue to the origin of pseudogap physics?

We believe that the key property is that in many of these materials the metal-insulator transition would, in the absence of frustration, be first order.  Consider elemental Si and Ge.  There are two ways to dope them, either by chemical substitution or by photodoping.  The results of photodoping are remarkably simple: at low temperatures the excited electrons and holes form excitons, which at high doping condense into metallic liquid droplets in an insulating background.\cite{EHD}  In contrast, with chemical substitution one carrier is rigidly locked to the lattice, and the transition involves unbinding of impurity-bound carriers.  
One might expect that nothing like electron-hole droplets could occur in a strongly correlated material, but one would be wrong. In La$_2$CuO$_{4+\delta}$, the oxygen dopants are intercalated between cuprate layers and remain mobile to fairly low temperatures, leading to macroscopic phase separation, with droplets of high-$T_c$ material in an AFM background, with doping-independent $T_c$ and Neel temperature $T_N$, and the highest $T_c$ of any La-cuprate.\cite{Jorg,BirgO2,Bian2,VHSrev}. Notably, adding Sr dopants to La$_2$CuO$_{4+\delta}$ rapidly degrades the droplets, leading to NPS with accompanying reduction of $T_c$ and doping-dependent $T_c$ and $T_N$ -- very similar to chemically doped Si.

We propose that cuprates follow an alternative route to strong correlation operative in many materials -- a frustrated first-order transition, leading to NPS.  We have shown that a model of the pseudogap as a phase of short-range AFM order can describe many properties of the pseudogap\cite{AIP,MBMB,Paper1}.

To understand such a frustrated order, one needs to answer four questions: 
What is the underlying order?
Why is it frustrated?
How does one describe the resulting disordered phase in the clean limit?
How does `dirt' (typically dopants) modify the disorder?
 
\subsection{Frustration and phase separation}

We find that starting from the AFM phase, the insulator-metal transition in cuprates is first order, and that undoped cuprates are incompressible.\cite{WhiSc}  Specifically, in a gapped material the Fermi energy has a discontinuity across the gap, which is reflected as a cusp minimum of the free energy vs density, with different curvatures on the two sides of the cusp.  If one curvature is concave, the free energy curve lies above a straight line connecting the free energy at half filling to that at any finite doping, meaning the material is unstable to phase separation.  In such a phase separation, a tangent construction shows that one stable end point is always at the cusp -- that is, the gapped phase is incompressible, and can only be doped by adding domains of a second phase.  In cuprates, we have proposed that this second phase is associated with the saddle point VHS, which is known to have a non-analytic free energy\cite{Rice}.  However, it was not clear which VHS-related instability is present in cuprates.\cite{SO8}  Below, we will propose a likely candidate.

When the dopant ions are essentially immobile in the lattice, electrons cannot phase separate by themselves without incurring a large Coulomb penalty.  This penalty can be eliminated by reducing the spatial scale of the phase separation to the nanoscale regime.\cite{RSMCondon2,VHSnps,VHSrev}  We propose that the resulting short-range order consists of a series of three stripe textures of an underlying AFM order, suggestive of charges confined on topological defects of the AFM order, the antiphase boundaries.  Figure~\ref{fig:6a}(a) illustrates  the experimentally observed $Q$-vectors of these textures. Our model of NPS stripes was presented in Refs.~\onlinecite{RSMstr,RSM2000}, with predicted $Q$-vectors shown in Fig.~\ref{fig:6a}(e)\cite{RSMstr}. Recall that the $S_2$ branch terminates at the AFM VHS, $x_{VHS}$, which we had taken as $x_{VHS}\sim$~0.25 in the La-cuprates, but which is substantially larger in the Bi-cuprates.

In Ref.~\onlinecite{Paper1} we presented evidence that the underlying short-range order in the pseudogap is AFM, even near the doping where the pseudogap terminates, and in the following subsection we present a definitive demonstration that this is the case.  In Ref.~\onlinecite{Paper3} we show that similar results follow from a generalized three-band model of cuprates.  The rest of the paper presents our current picture of the resulting disordered phase.  Since we believe the main source of frustration is intrinsic, we will only briefly discuss the role of dopant disorder.

\subsection{AFM nesting of the CDW}

In Fig.~\ref{fig:6a}(a), branches $S_1$ and $S_2$ display an approximate symmetry which is captured in our model, Fig.~\ref{fig:6a}(e).  Within the NPS model, this symmetry arises since the $S_1$ branch starts in the pure AFM phase, and becomes doped by adding 2-Cu-wide charge stripes, while the $S_2$ branch starts at $x_{VHS}$ and loses charge by adding 2-Cu-wide AFM stripes. The symmetry is only approximate due to differences in interaction between the stripes. It is widely recognized that stripes in the $S_1$ branch have simple repulsive interactions (confirmed by DFT calculations\cite{YUBOI}).

Here we focus on the $S_2$ branch. Since the incommensurate $Q$-vector of the $S_2$-branch has the opposite doping dependence from that of the $S_1$ stripes, it is a candidate for Fermi-surface nesting, leading to its identification as a CDW. A debate has arisen as to whether the nesting is Fermi surface nesting\cite{EHud1} or `hot spot' nesting\cite{Comin}.  While nesting of the NM Fermi surface has a qualitatively similar doping dependence, it cannot quantitatively fit the $S_2$ line\cite{EHud2,Gutzcharge}, as we confirm in Fig.~\ref{fig:6a}(a), thick violet dot-dashed line.  Instead, here we show that the nesting is associated with the Fermi surfaces of the lower Hubbard band (LHB) of the AFM phase.

Since the AFM Fermi surface consists of electron and hole pockets similar to those found in electron-doped cuprates, there are several possibilities for nesting, discussed in SM III.A.  We find that a form of hot-spot nesting between two adjacent hole pockets, green arrow in Fig.~\ref{fig:6a}(b), provides a good description of the full $S_2$ curve -- violet dashed line in Fig.~\ref{fig:6a}(a).  Frame (c) illustrates the result of simultaneously nesting along $x$- and $y$-axes.  If we recall from Ref.~\onlinecite{Paper1} that the red parts of the curves are shadow bands with low intensity, the resulting Fermi surface is the star-shaped violet curve, the shape proposed for the QO Fermi surface\cite{Seba}.  In frame (d) we illustrate the evolution of the Fermi surface area as doping increases and the magnetic gap is reduced.  Note that this clears up another puzzle: why do the QOs always have the same Fermi surface area, except in Hg cuprates\cite{GrevenQO}?  Because most represent the $P_c=4$ $S_0$ phase, while Hg cuprates are on the $S_2$ line.  A simple $6\times 6$ model was able to reproduce the correct Fermi surface shape for commensurate $(P_c=6)\times (P_c=6)$ crossed stripes\cite{Gutzcharge}.  

Similar calculations for incommensurate stripes will be harder, as they seem to be a mixture of two adjacent commensurate periodicities, $P_c$ = $N$ and $N+2$ (since even-width stripes have lower energy)\cite{YUBOI}.  Experimentally, such a mixture is found to arise by bifurcations of individual stripes, with each bifurcation leading to a vortex in the magnetic domain wall
\cite{Nematic}.  That is, the incommensurate stripe phase can be accomplished by generating vortices along the domain walls -- i.e., {\it by inducing an Ising-XY transition}.  This could explain the abrupt disappearance of stripes before the pseudogap collapse, since in the transition the topological defects change from domain walls to point vortices.

This is a very significant result.  First, it is a {\it nesting of the AFM Fermi surfaces}, proving that the CDW stripes are excitations of an underlying AFM order, consistent with Ref.~\onlinecite{Paper1}.  Secondly, we find that there are two symmetry related nesting vectors, $Q_x$ and $S_x$, Fig.~\ref{fig:6a}(b), which satisfy $Q_x+S_x=2\pi/a$, consistent with experiment.\cite{EHud1}  Finally, we will demonstrate that the accompanying C-I transition is a form of Mott-Slater transition.

\subsection{Nanoscale phase separation}
We have developed an NPS model of cuprates\cite{RSMstr,RSM2000} to explain neutron scattering data on the stripe phases\cite{Tranq1} of branch $S_1$ in Fig.~\ref{fig:6a}(a). Initial many-body calculations discussed in SM II were used to simplify the calculations. Notably, the same model can also describe branches $S_0$ and $S_2$, Fig.~\ref{fig:6a}(e).  Here, we briefly review the model and its applications to cuprates, and in later sections we introduce new results.  

The original model assumed two components with a first-order transition between them.  The corresponding free energy minima are associated with AFM order at $x=0$ and another, unspecified order at the VHS doping, $x_{VHS}$, whose nature is discussed further below.  The stripe phase is an intimate mix of these two phases which evolves smoothly from one to the other with doping. Starting from $x=0$, topological defects -- antiphase boundaries acting as domain walls -- are dressed with added holes, leading to an average doping on the charged stripes of $x=0.25$ holes, close to $x_{VHS}$ for La-cuprates. The average doping increases by adding more charge stripes, which repel each other forming a uniform stripe array.  It is these dressed domain walls which keep the AFM order from becoming long-ranged.

For this particular model, an electrostatic version of St. Venant's principle can be formulated.  Electrostatics is based not on the microscopic, rapidly-varying electric field but on a smoothly varying macroscopic average field.  Since this average must be over a volume large compared to a unit cell, these striped phases produce no excess Coulomb fields.  

Results of our model include (1) there is a natural maximum nesting vector $Q$ associated with the narrowest stripe repeat unit, and (2) beyond the doping of maximal $Q$, the NPS model predicts a second branch of stripe phase, where the AFM phase forms domain walls while the charge phase gets wider.  Since the stripes are regions of the two end phases, their minimum width would each require 2 Cu atoms (as now confirmed by DFT calculations\cite{YUBOI}), leading to $Q_{min}=P_c=4a$.   
(3) While this provides a good model for the first branch, the neutron data showed no sign of the second branch, but instead leveled off beyond $x_{VHS}/2$, the doping of the $P_c=4$ stripes, leading to our suggestion that the minimal-width stripe phase is particularly stable, and persists over a finite doping range, consistent with branch $S_0$.  (4)
Later, when evidence for CDWs was found, they fell into two groups, one with doping independent $Q$ consistent with $P_c=4$, and a second with a $Q$ that decreased with doping, similar to branch $S_2$.  Experimentally, the minimum $P_c$ is found to be $4a$, consistent with the NPS model, while DFT calculations find that the $P_c=4$ phase is the lowest energy stripe phase in YBa$_2$Cu$_3$O$_7$ (YBCO$_7$)\cite{YUBOI}.  

\begin{figure}
\leavevmode
\rotatebox{0}{\scalebox{.5}{\includegraphics{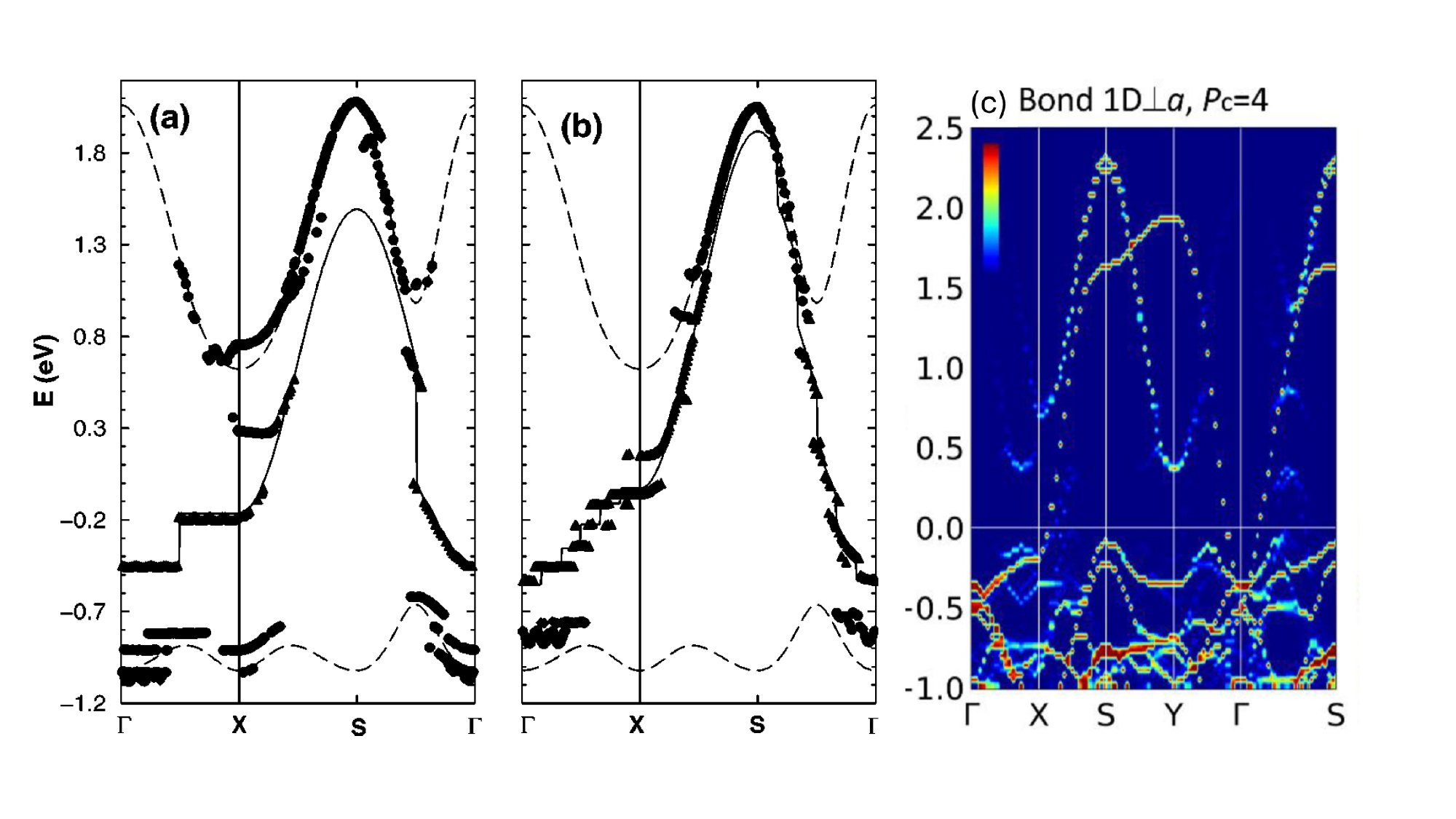}}}
\vskip0.5cm
\caption{(Color online)
{Stripe dispersions, for $(m,n)$ stripes, where the magnetic stripes are $m$ Cu wide and the charge stripes are $n$ Cu wide}.  (a) (6,2) (type $S_1$) stripes. (b) (2,6) (type $S_2$) stripes, and (c) (2,2) stripes.  In (a) and (b) symbols represent the stripe dispersion while for comparison the dashed lines represent the half-filled AFM insulator and the solid line represents the dispersion of a single charged stripe.  Note that in (c), the result of a DFT calculation, all bands are displayed, but the Cu-O AB band is readily distinguished.  Note that in units of $\pi/a$, $X=(1,0)$, 
$Y=(0,1)$, and $S=(1,1)$. Frames (a) and (b) are from Ref.~\onlinecite{RSMstr} and frame (c) is from Ref.~\onlinecite{YUBOII}.
}
\label{fig:11}
\end{figure}

To better understand the stripe phases we carried out Hartree-Fock calculations on stripe arrays, including their effect on the underlying AFM phase.\cite{RSMstr,WhiSc}  For a 1D stripe array, the unfolded dispersion was found to clearly resemble the underlying AFM dispersion, but with extra minigaps in the direction perpendicular to the stripes, Figs.~\ref{fig:11}(a,b).  Such stripe-related minigaps have now been found in DFT calculations of cuprates\cite{YUBOII}, Fig.~\ref{fig:11}(c), and of a related stripe phase in the infinite-layer nickelates.\cite{Nick}  Moreover, while the average charge density was set at 0.25 on each charge stripe and 0 on each magnetic stripe, the charge distribution relaxed so the average charge modulation was much smaller, consistent with DFT calculations\cite{YUBOI}. 
This latter point is in direct analogy to the spread of the magnetic fields off of the vortex cores in a type II superconductor, so that the actual field modulation is never larger than the lower critical field $H_{c1}$, whereas  superconductivity can persist to the much higher field $H_{c2}$.

While this model describes many aspects of the textures found in the pseudogap, there were also a number of problems.  
(i) If the stripes are associated with nesting the NM Fermi surface, the $P_c=4$ stripe phase would fall at electron doping, inconsistent with experiment.\cite{Seba,Ramshaw}  As we demonstrated above, these stripes actually involve nesting with the underlying AFM Fermi surface.
(ii) The reason for the stability of the $S_0$ branch remained unclear.  In the next Subsection, we provide a clear explanation for this three-branch spectrum.
(iii) The particular phase associated with the VHS endpoint was not specified.  In Section III.A we show that recent experiments suggest a solution to this problem.

\subsection{More is different: the electronic eutectoid model}
To understand the special role of the $P_c=4$ phase, we return to alloy theory.  We begin with a brief review of eutectics, a topic unfamiliar to many condensed matter researchers.  

Anderson's encomium\cite{PWA}, often applied to pseudogap physics, can equally well be applied to the physics of eutectics/eutectoids.\cite{steel}  In many binary alloys, instead of simple two-phase coexistence regimes, a third phase intervenes.  An intimate mix of the two end phases can lower their melting point to the extent that the resulting liquid has lower energy than the coexistence tie-line.  As temperature is lowered, the liquid becomes less stable, ending at a unique eutectic point at fixed composition and temperature.  Below this eutectic temperature, a new phase is formed as an intimate mix of the two end phases -- typically lamellar.  For cooling away from the eutectic doping, the system phase separates into the nearer end point and the eutectic phase.

Even more similar to the cuprates is the eutectoid phase, where the liquid phase is replaced by a solid phase, while the other phases remain the same.  The best known example is iron with low concentrations of carbon.  Immediately below the austenite eutectoid point, an intimate lamellar mix of the two main phases, ferrite and cementite (Fe$_3$C), forms, called pearlite.  Away from the eutectoid point mixtures of pearlite and either ferrite (for lower carbon concentrations) or cementite (at higher concentrations) form.  

Thus, cuprates can be considered the NPS limit of a eutectoid system, with branch $S_1$ acting as the hypo-eutectoid, branch $S_2$ the hyper-eutectoid, and the $S_0$ state being the intimate mix that forms below the eutectoid point. We note that already in pearlite, the mesoscopic layering of ferrite and cementite leads to exotic physical properties.  The name pearlite comes from its opalescent optical properties, reminiscent of pearls and opals, and it is now considered a natural photonic crystal.  Table 2 summarizes our comparison of the cuprate and Fe-C systems.

   \begin{center}
Table 2: Comparison of Iron Eutectoid and Cuprate Pseudogap\\
\begin{tabular}{ |c|c|c|c| } 
 \hline
Property & Fe system & cuprate system & Fe eutectic   \\
 \hline
 {\it Slow cool} &&&\\
 \hline
 Liquid-like phase & austenite & mobile oxygen phase & melt \\  
 Eutectoid & pearlite & $P_c=4$ phase & ledeburite \\
 Low-$x$ phase & ferrite & AFM phase & pearlite   \\ 
 High-$x$ phase & cementite & LTT phase  & cementite \\ 
 \hline
 {\it Fast cool} &&&\\
 \hline
 Phase & -- & --  & bainite  \\
 \hline
 {\it Quench} &&&\\
 \hline
 Phase & -- & pseudogap  & martensite  \\  
 Proliferation of & -- & domain walls & dislocations  \\
 \hline
\end{tabular}
\end{center}

Tying this picture more to NPS, it is found that near the eutectoid transition temperature the carbon has high mobility, so that the two constituent phases of pearlite are formed by carbon diffusion, while the alloy can easily fall out of equilibrium by increasing the cooling rate.  This is best exemplified in the Fe-C system at a slightly higher C content, where a true eutectic forms at a higher temperature than the eutectoid temperature (fourth column of Table 2).  The eutectic phase that forms below the melt is called ledeburite, and when slowly cooled it decomposes at the eutectoid temperature to a mixture of pearlite and cementite.  However, the fate of ledeburite is highly sensitive to cooling rate, and as the cooling rate is increased two intermediate phases can be formed, bainite and martensite.  Whereas pearlite forms by C diffusion, martensite cools so rapidly that diffusion is frustrated and the phase forms by a diffusionless shear process.  Cooled at intermediate rates, bainite forms by a mixture of diffusion and shear.  Indeed, there are several subvariants of bainite, depending on the cooling procedure.\cite{Bainites}   Faster cooling leads to more intimate mixtures of the component phases.  Of particular interest is the martensite phase, which has a high dislocation density suggestive of Kibble-Zurek physics.\cite{KZ1,KZ2}  

Frustration can also be induced by increasing the number of atomic constituents, as in eutectic high-entropy alloys, where lamellar widths under 100~nm can be found.\cite{EHEA}  Thus, increasing frustration systematically pushes eutectics/eutectoids toward the atomic-scale limits of NPS.  Similar effects are found in spinoidal systems\cite{ChargeSpinodal}, where adding charge modulations tends to quench the modulation periodicity.

In cuprates, substituting Sr or Ba for La leads to very low-mobility ions and NPS, while, as noted above, when oxygen is intercalated into La$_2$CuO$_{4+\delta}$, the oxygens are mobile at room temperature, leading to macroscopic phase separation into undoped AFM and optimally doped superconducting domains\cite{Jorg,BirgO2,Bian2}.  Finally, we note that the related material Sr$_2$RuO$_4$ can form eutectic phases when mixed with excess Ru\cite{Sr2RuO4A} or with Sr$_3$Ru$_2$O$_7$\cite{Sr2RuO4B}, and it has been suggested that the $T_c=3K$ high-$T_c$ phase forms on Sr$_2$RuO$_4$-Ru domain walls\cite{Sr2RuO4A}. 

In summary, in the Fe-C phase diagram, the pearlite eutectoid phase is an intimate mix of the two end phases ferrite and cementite.  At the same time, it is the high doping end of a two-phase coexistence regime of ferrite and pearlite, and the low doping end of a second two-phase coexistence regime of pearlite and cementite.  In the cuprate pseudogap, the $P_c=4$ phase can be thought of as the most intimate mix of the AFM and VHS end phases.  At the same time, it constitutes the high-doping end of the stripe NPS regime $S_1$ (with the narrowest stripe repeat) and the low-doping end of the CDW NPS regime $S_2$ (with the smallest star-shaped Fermi surface).

The $P_c=4$ phase has all the characteristics of the NPS limit of the eutectic, as found in Fe-C binary compounds, except in the limit of atomic interlamellar distances.  Thus, the average interlamellar distance in pearlite is $\sim$320~nm\cite{Pearlite}, which can be reduced to under 100~nm in the frustrated eutectic high entropy alloys, and to $\sim$3~nm in cuprates.  Moreover, a eutectoid is a phase of fixed composition,thereby explaining why the $Q$-vector and Fermi surface area are nearly invariant over a wide doping range.  The above stripe models\cite{RSM2000,RSMstr} provide a good model for 1D $P_c-4$ stripes at 1/8 doping.  Finally, we anticipate that near 1/8th doping the role of oxygen density fluctuations will be different.  Deep inside the two-phase regime, the fluctuations will be too small to move the local hole density to one of the stable end phases, so the holes will adopt the average local density, leading to the well-known local patches and gap maps. However, near 1/8th hole doping the fluctuations can move the local density to or beyond 1/8th.  Then, the local hole density can greatly expand the area of sample at or very close to the eutectoid, thereby explaining the percolation-like behavior observed experimentally.\cite{Ramshaw}

\section{What is the VHS end phase?}
\subsection{Origin of $d$-form factor density wave}

A major issue is, what is the nature of the phase that is stabilized near $x_{VHS}$?  Since the high DOS near $x_{VHS}$ can drive a large number of instabilities\cite{SO8}, it is not clear that this phase is the same in all cuprates. Excluding the macroscopic phase separation of La$_2$CuO$_{4+\delta}$, in other La-cuprates it was suggested that the low-temperature tetragonal (LTT) phase played this role, as it was known to involve a splitting of the VHS singularity.\cite{LTT1,LTT2,LTT3}  However, the LTT phase seemed to be stable only in a limited doping range of La$_{2-x}$Ba$_x$CuO$_4$, with only fluctuations in other La-cuprates, and with no analog in other cuprate families, so this idea was largely set aside.  Another possibility is that the identity of this phase is obscured by its interactions with superconductivity.

Recent experimental developments call for a reexamination of this issue.  First, it was found that the stripe phases in the (La.Ba) cuprates have an unusual stacking along the $c$-axis, leading to a 4-CuO$_2$-layer thick unit cell.  This has been interpreted as a locking of the stripes to an underlying LTT order, as follows.  The LTT phase involves octahedral tilts along the $x$-directed Cu-O-Cu bond in one layer, and along the $y$-directed bond in the next.  It was postulated that the charge stripes aligned with these tilts, so that the stripes in one layer ran along $x$, and in the next along $y$.  The second period doubling along $c$ was interpreted as a Coulomb effect.  To reduce Coulomb repulsion, the $x$-directed stripes on the first and third layers are offset by half a unit cell.  This model also provided the clearest explanation of why the stripes run along $x$ and $y$ rather than diagonally.   While this model is now widely accepted for the (La,Ba)-cuprates, it raises a number of issues.  First, the stripe phase in the (La,Ba)-cuprates extends over a much wider doping range than the LTT phase, and indeed stripes in most cuprates align along the $x$ and $y$ axes, even though they lack an LTT-like phase.  An even more serious issue arises involving the QOs found in many cuprates.  The small Fermi surface can readily be understood in a model of crossed (2D) stripes, but other experiments are more consistent with 1D stripes.  A possible solution for this is that the stripes in any layer are 1D, but the stripes in different layers alternate in directions just as found in the (La,Ba)-stripes\cite{crossed,Ramscrossed,Tranqcrossed}.  But again, this would strongly suggest that an LTT-like phase is a universal feature of cuprates.

Recent STM experiments provide much more direct evidence.  A key signature of the role of the VHS in the LTT phase is a symmetry lowering, in which the two CuO$_2$ planar oxygens become inequivalent.\cite{LTT1,LTT2,LTT3}.  The STM experiments identify this splitting as a key signature of a new phase found in all cuprate stripe phases studied (Bi- and oxychloride cuprates), labeling them as $d$-form factor density waves (dffdw), and proposing that they are a universal feature of cuprates.\cite{d-form,Oorbital}

\subsection{Modeling the dffdw}

\subsubsection{LTT-phase}

We propose that the NPS end phase at $x_{VHS}$ is an LTT-like phase.  The LTT phase in La-cuprates, consisting of a particular tilting pattern of the CuO$_6$ octahedra, has been studied in some detail.  Long-range LTT order is found only in special cases -- La$_{2-x}$Ba$_x$CuO$_4$ in a limited doping range and some materials with partial replacement of La by rare earth elements.  The LTT phase is stabilized by splitting the VHS, but this splitting is surprisingly small,\cite{Buchner}  which is not uncommon for dynamical Jahn-Teller (JT) effects.  

While the LTT phase in LSCO involves octahedral tilts, we explore a simpler purely electronic model which displays similar physics.  This electronic LTT phase has a dispersion that differs from the conventional nonmagnetic dispersion only in that the nearest-neighbor hopping parameter is different along $x$- and $y$-axes, $t_{x,y}=t(1\pm \delta t)$.  Below, we analyze this model for fixed $\delta t=0.1$, and for the most part for La-cuprates, $t'/t=-0.13$.\cite{Paper1}  

\begin{figure}
\leavevmode
\rotatebox{0}{\scalebox{0.50}{\includegraphics{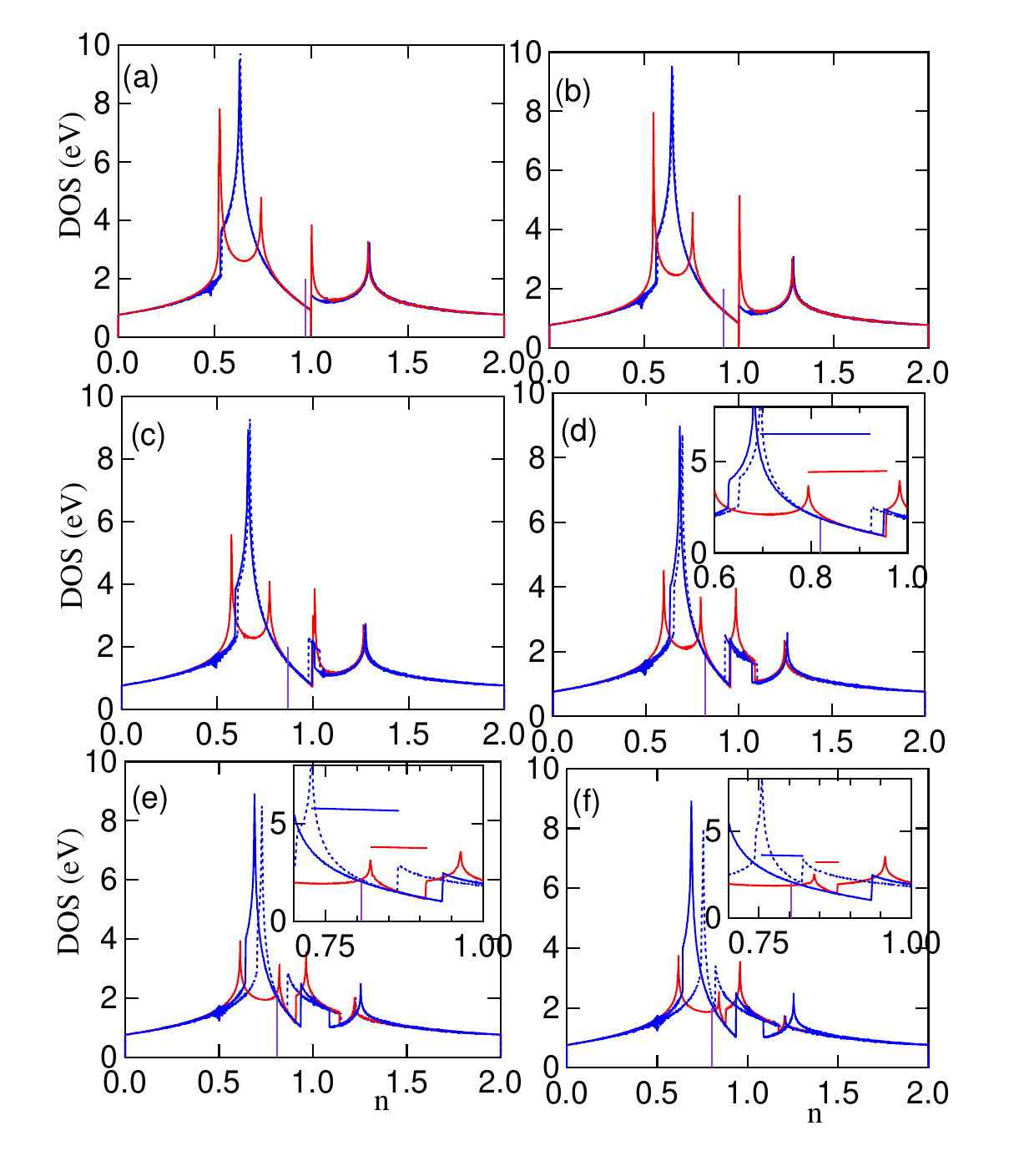}}}
\vskip0.5cm
\caption{(Color online)
{\bf Model of the nematic phase as an AFM-LTT order}.  (a-f) Comparing the static LTT (red) and NM (blue) DOS for AFM gap $\Delta$ = 0.25 (a), 0.20 (b), 0.15 (c), 0.10 (d), 0.05 (e), and 0.02 eV (f).    The vertical violet lines indicate the self-consistent density in the LTT phase. Inserts show blowup of the closing AFM gaps (horizontal lines).
}
\label{fig:B4}
\end{figure}

\begin{figure}
\leavevmode
\rotatebox{0}{\scalebox{0.70}{\includegraphics{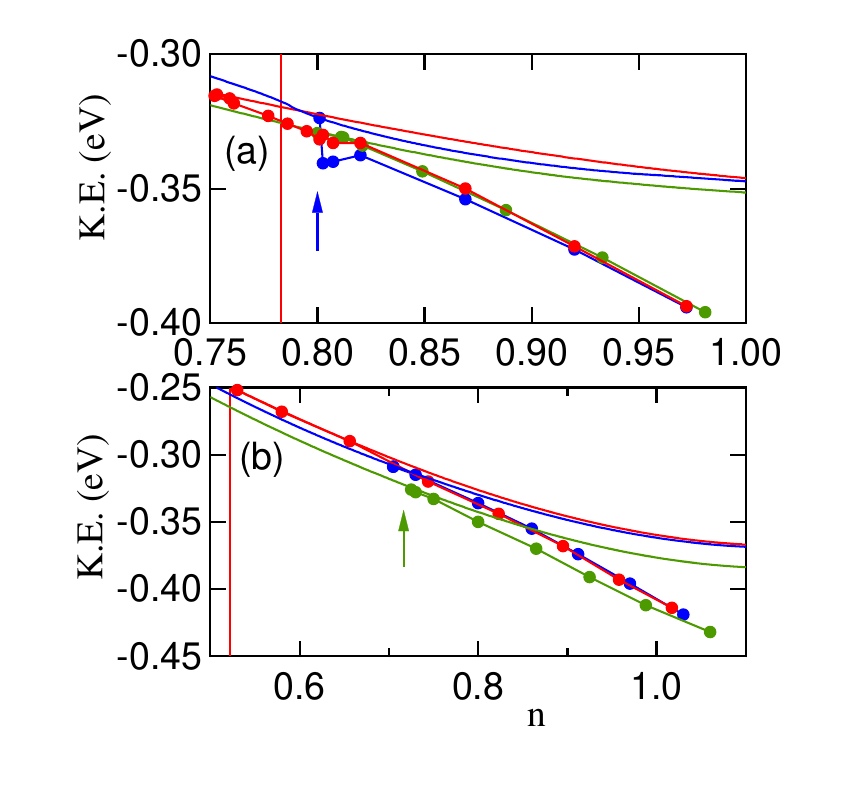}}}
\vskip0.5cm
\caption{
{\bf Model of the nematic phase as an AFM-LTT order}.  Kinetic energy  (K.E.) vs doping $x$ in pure AFM phase (blue) vs combined AFM + LTT (red) or AFM + LTO (green) phases, for $t'/t$ = (a) -0.13 (La-cuprates) or (b) -0.25 (Hg-cuprates).  Curves without dots represent $S_d=0$ phases; vertical lines = doping of bare VHSs; arrows = termination of lowest energy AFMs. 
}
\label{fig:B4b}
\end{figure}

We focus on one issue: neglecting stripe order, is AFM order more stable in the LTT or NM phase, where the latter is defined to be the phase with $\delta t=0$.  One must be careful in comparing phases: in all phases, $S_d$ is found self consistently, and one compares phases with the same electron number $n$, which means the values of $S_d$ are different in the different phases.
Figure~\ref{fig:B4} compares the DOS for the AFM phase (blue) with a combined AFM plus LTT phase (red), with each frame corresponding to a different value of $S_d$ for the mixed AFM-LTT phase. The dashed blue curve represents the self-consistent pure AFM phase with the same value of $S_d$, while the solid blue curve represents the AFM phase at the same doping (indicated by the vertical violet line), and hence with a slightly larger value of $S_d$.  It can be seen that the LTT phase is stabilized by the splitting of the AFM VHS.  The AFM gap, which is calculated self-consistently\cite{Paper1}, is not very sensitive to the LTT vs NM order, except when the AFM gap is small.  In the NM phase, the gap collapses when the bottom of the upper magnetic band is close to the VHS peak, whereas in the LTT phase the collapse is associated with the upper VHS.  Thus, the AFM gaps in the NM and LTT phases remain close until frame d, when the gap encounters the upper VHS in the LTT phase, while the NM gap continues shifting to lower energies in frames e,f.  To estimate the stability of the various phases, in Fig.~\ref{fig:B4b}(a), we plot the kinetic energy (K.E.) of the pure AFM (red), the AFM+LTT (blue), and the AFM+LTO (green) phases.  The latter is discussed in the next subsection.  Over the full doping range of the AFM+LTT phase, it has the lowest kinetic energy, with the lowering increasing as the VHS is approached. However, the AFM phase coexisting with the LTT phase collapses near the upper VHS (blue arrow), and for larger hole doping (smaller $n$) the blue curve represents the pure LTT phase, which has higher energy than the pure AFM phase.

\subsubsection{LTT-LTO phase transition}
Above, we compared the LTT-phase to the NM phase.  However, in La-cuprates there are several phases underlying the AFM phase which are closely related to the LTT-phase, chief among them the low-temperature orthorhombic (LTO) and high-temperature tetragonal (HTT).  Nominally, the LTT and LTO phases differ in the direction of their octahedral tilts, while the HTT is untilted, but we proposed that the LTT represents a static Jahn-Teller (JT) phase, while the other two are different dynamic JT phases, with the tilt angle adjusted by hopping between 2 (LTO)  or 4 (HTT) equivalent LTT-type tilts.\cite{VHSrev}  It has now been demonstrated that indeed, in the LTO and HTT phases the local tilts are LTT-like, but this is usually interpreted in terms of a static order-disorder transition.\cite{LTTpatch,EgBill,Haskel,Bozin}

\begin{figure}
\leavevmode
\rotatebox{0}{\scalebox{0.50}{\includegraphics{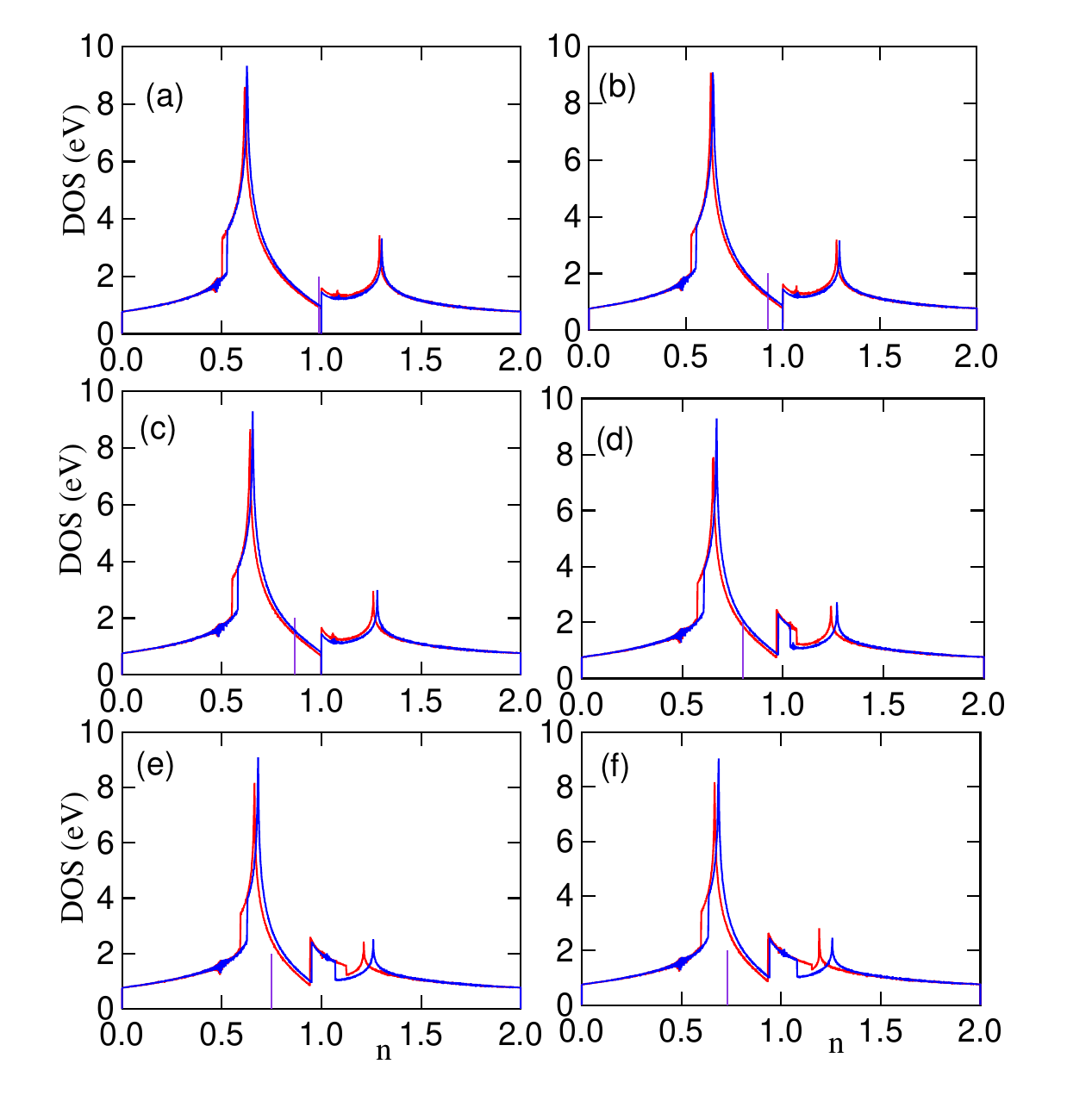}}}
\vskip0.5cm
\caption{
{\bf Model of the nematic phase as a dynamic AFM-LTO order}.  (a-f) Comparing the dynamic LTT, or LTO (red) and NM (blue) DOS for AFM gap $\Delta$ = 0.25 (a), 0.20 (b), 0.15 (c), 0.10 (d), 0.05 (e), and 0.02 eV (f). 
}
\label{fig:B5}
\end{figure}

Here we show that our LTT model can be extended to an LTO model.   In our earlier work\cite{RSM8C}, we postulated that the octahedral tilts adjusted on an atomic scale, leading to an orbital antiferromagnet (OAF)\cite{Schulz}.  However, recent experiments find local patches of x- or y-type LTT tilts\cite{LTTpatch}, raising the possibility of collective rotation over the patch.  In this case, we can model the dynamic nuclear distortions by making $\delta t$ imaginary, consistent with Ref.~\onlinecite{HalpRice}.  Figure~\ref{fig:B5} compares the DOS of the pure AFM phase (blue) with the DOS of the AFM+LTO phase (red), where for simplicity we refer to the dynamic LTT phase as the LTO phase, even though technically it is closer to the HTT phase.\cite{RSM8C}  Remarkably, the DOSs for both LTO and NM phases look quite similar, with no obvious splitting of the VHS in the LTO phase.  This can be understood by looking at the energy of the AFM state.  For pure AFM, AFM+LTT, and AFM+LTO phases, the magnetic bands have the same form of dispersion, $E_{\pm}=\epsilon_+\pm\sqrt{\Delta^2+|\epsilon_-|^2}$, with only the $\epsilon_-$-term changing.  For the pure AFM, $\epsilon_-=-2t(c_x+c_y)\equiv\epsilon_{0}$, while for the AFM+LTT, $\epsilon_-=2(t_xc+x+t_yc_y)$, where $t_{x/y}=t(1\pm\delta t)$.  Since $t_x\ne t_y$, the VHS is split.  In the AFM+LTO phase, $\delta t\rightarrow i\delta t$, so $|\epsilon_-|^2 = |\epsilon_{0}|^2+\Delta^{'2}$. with $\Delta'=2t\delta t(c_x-c_y)$, and the energy is identical to that of the pure AFM, except with a larger gap, $\Delta^2\rightarrow\Delta^2+\Delta^{'2}$. In other words, the LTO-phase is a {\it hidden order}, lowering the energy of the AFM phase without clearly showing up in the DOS.  Despite this, we find that the AFM+LTO phase terminates at nearly the same doping as the AFM+LTT phase. In Fig.~\ref{fig:B4b}(a) we plot the K.E. of the AFM+LTO phase in green.  We find that the AFM+LTT phase has lowest energy as long as the AFM component is finite, but it terminates at the blue arrow for both mixed phases, and for larger hole doping the pure LTO phase has lowest energy.  That is, the pseudogap appears to terminate at the upper-VHS of the LTT phase.  While this is suggestive of a result of Ref.~\onlinecite{Paper1}, it must be confirmed by self-consistent calculations of the LTT phase. Recently, it was found that in La$_{1.88}$Sr$_{0.12}$CuO$_4$ static and dynamic stripes coexist, both with the same tilt angle, associated with anisotropic values of $t'/t$.\cite{tpanisot}

The present results offer striking support for the prediction that the cuprate LTO structure is actually a dynamic LTT.\cite{VHSrev} Note that the orbital AFM phase should readily interact with the chiral phonons that were recently observed in the La-cuprates.\cite{Taill1}  In Fig.~\ref{fig:B4b}(b) we plot the corresponding K.E.s for Hg-cuprates, $t'/t=-0.25$.  In this case the AFM+LTO phase has lowest energy as long as the AFM order is finite, perhaps explaining why no LTT-like phase has been observed in these cuprates.

We note the following: 
(1) Ref.~\onlinecite{d-form} found two superimposed anomalies: the dffdw at a $Q$-vector consistent with the $S_2$-stripe branch, and a nematic phase at $Q$=0, where the $x$- and $y$-Bragg peaks are inequivalent.  The AFM-LTT-phase in Fig.~\ref{fig:B4} has characteristic properties matching the observed nematic phase, while the corresponding $S_2$-stripe phase, composed of alternating AFM and LTT stripes, is a dffdw. Due to the patch map structure of the cuprates, both phases can be found in a single sample.\cite{gapmap}

(2) In Ref.~\onlinecite{Paper1}, we showed that the pseudogap collapse resembles the collapse of the AFM phase to a NM phase.  Since we have now found that the end phase is an LTT-phase rather than NM, we need to understand how the picture of pseudogap collapse is modified.  This is illustrated in Fig.~\ref{fig:B4}(b-h).  The AFM phase still terminates near the doping of the NM VHS, but there is now an interference with one of the LTT VHSs.  This has the potential of explaining two puzzles in the earlier results.  First, the transition was found to occur when the Fermi level enters the region of rising (VHS related) DOS, and to be first-order for small-$t'$ cuprates.  Yet a full, logarithmically-diverging DOS was found experimentally.  From Fig.~\ref{fig:B4}(b), we find that the collapse occurs when the AFM peak intersects the higher-energy LTT-VHS, Thus, the observed heat-capacity peak could be associated with the lower LTT-VHS (at higher hole doping).  Secondly, the higher LTT-VHS moves to lower doping with increasing $\delta t$, which could explain why the experimental transition from small to large Fermi surface seems to occur at a lower doping than predicted, and the deviation gets worse as $t'$ gets larger.  For example, in Fig.~\ref{fig:B5}(f), the LTO VHS falls near $x_{VHS}=0.5$, as expected for $t'/t=-0.26$, whereas in the LTT phase the upper VHS falls at $x=0.27$, close to the $x=0.32$ found experimentally, Fig.~4(c) of Ref.~\onlinecite{Paper1}.  However, the model calculation assumed a $t$ modulation of 10\%, which leads to an energy splitting of about 160~meV, larger than the experimental average $\sim$50~meV in Bi-2212,\cite{Oorbital} suggesting that the real splitting is closer to 3\%.

(3) In Ref.~\onlinecite{Paper3} we show that a similar LTT-phase arises in a generalized 3-band model.

(4) The doping-dependent LTO-LTT crossover in Fig.~\ref{fig:B4b}(a) is in good agreement with the experimental LTO-LTT crossover in the La-cuprates.  Moreover, the LTT-phase should persist beyond the pseudogap collapse, and the associated CDW instability may have already been observed.\cite{reent}

\subsubsection{Possible relation to thermal Hall effect}

Finally, the pseudogap was recently found to host a thermal Hall effect (THE), in which the motion of the phonons is modified by a magnetic field.\cite{TaillThPh}  This effect is believed to be associated with chiral phonons,\cite{Taill1} which can be allowed if the crystal symmetry is locally lowered\cite{THE1}.  Indeed, evidence for such lower symmetry has been found in several cuprates.\cite{nematic0,chiralG,chiralN,Tranq_chir}  Many of these lower symmetry phases are compatible with the JT phases discussed above.

We note that similar dynamic-JT-like phenomena arise in titanates.  The tilt-mode phases in SrTiO$_3$ are particularly similar\cite{STOoct}, but the ferroelectric distortions in BaTiO$_3$ also display similar phenomena.\cite{HelPhys,VHSrev}  Notably, it was the study of these modes that led M\"uller to look for similar phenomena in cuprates.  

In La$_{2-x}$Ba$_x$CuO$_4$, it was found that the LTO phase is composed of a coherent superposition of LTT patches, with each patch having a size of $\sim$1~nm.\cite{LTTpatch}  This size is similar to that of the gap patches seen in scanning tunneling microscopy (STM) studies, which control the local size of the AFM and superconducting gaps.  Since the LTT-LTO-HTT system involves a form of NPS, it was hypothesized to be related to the AFM-superconducting properties of cuprates, but a direct connection was not found. The present results now clarify this connection.

While dynamic Jahn-Teller transitions are mainly studied in molecules\cite{Na3} or at defects in solids, the La-cuprates would constitute a lattice of Jahn-Teller sites, and edge-sharing octahedra would couple the dynamic oscillations across different octahedra.\cite{RSM8C}  Similar collective oscillations were proposed in C$_{60}$ crystals.\cite{Berryonic1,Berryonic2}  These would seem to be a form of time crystals.\cite{Berryonic4,timerev} Recently there has been an experimental observation of a bulk dJT material (with the title, ``And Yet It Moves")\cite{andyet}.  We note that while the ground state properties are sensitive to the heights of tunneling barriers, above the transition {\it the soft phonons should be chiral} and locally resemble `time-molecules' such as Na$_3$\cite{Na3}.

As noted above, experiments clearly find that the LTO (and HTT) phases consist locally of LTT distorted domains.  However, theories diverge depending on whether the disorder is static (order-disorder transition) or dynamic (dJT transition).  According to Fig.~\ref{fig:6a}(a), the cuprates offer a new possibility.  By comparing STM and RIXS results, it is seen that due to competition between $S_0$ and $S_2$ stripes, when one is static the other is frustrated and remains metastable (dynamic), similar to the ferroelectric transition in SrTiO$_3$.  That is, order-disorder and dJT domains can coexist in the sample.  This can be restated in a suggestive way.  It has been found that in some La-cuprates the soft phonon mode is chiral.\cite {TaillThPh}  While it is not clear when the phonon finally softens whether the phase variable becomes static, that certainly is a real possibility, which can result in an order-disorder transition. However, it is also possible that the transition is frustrated, leaving behind a system with an ultralow phonon frequency.  Recent observation of a superconducting diode effect in Bi-cuprates could provide further evidence of a dynamical effect\cite{SCdiode}.

One final complication is that the underlying molecular JT model, denoted $E_g\times e_g$, can display chaotic solutions\cite{JTchaos1,JTchaos2} -- the condensed matter form of this could presumably be described as a time glass.  Alternatively, there is electron microscope evidence that, as a solid-state transition, the LTT phase never reaches equilibrium.\cite{Horibe}

\section{Superconductivity}

To put our results in context, in SM IV.E we briefly summarize the current status of modeling cuprate superconductivity.  In brief, there is much room for improvement, few models have been ruled out, and a better understanding of the `normal phase' is an important desideratum.  

\begin{figure}
\leavevmode
\rotatebox{0}{\scalebox{1.0}{\includegraphics{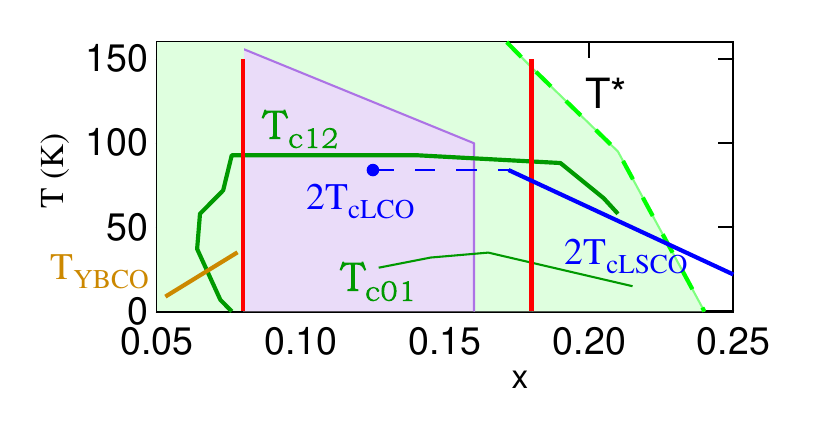}}}
\vskip0.5cm
\caption{
{\bf Superconductivity in the peudogap}  (green shaded region below $T^*$).  Purple shaded region = approximate domain of $P_c=4$ phase, as discussed in SM Fig.~S1.
The green curves labeled $T_{c(n-1)n}$ are the superconducting $T_c$s of Bi22(n-1)n, $n=1,2$\cite{Vishik,EHud3}.  The red vertical lines indicate the dopings at which the QO effective mass diverges\cite{Ramshaw}, presumably associated with percolation of the $P_c=4$ stripe phase.  Brown and blue curves represent $T_c$ for YBCO\cite{UDYBCO} and LSCO\cite{Boz1}, respectively.  Blue dot is for $T_c$ of oxygen-doped La$_2$CuO$_4$.
}
\label{fig:6c}
\end{figure}

Our model of the pseudogap provides a striking qualitative picture of cuprate superconductivity.  From the two phase coexistence in La$_2$CuO$_{4+\delta}$ we learn that the superconductivity is mainly associated with the eutectoid phase, which in other cuprates becomes the $P_c=4$ stripe phase.  Further evidence of this is found in Figure~\ref{fig:6c}, which is an abbreviated version of the cuprate phase diagram, Figure~\ref{fig:6b} below, focusing on the pseudogap (green shading) and the range where the $P_c=4$ phase dominates (purple shading), while the vertical red lines indicate the percolation limits of the $P_c=4$ QOs.\cite{Ramshaw}  Here we superimpose the superconducting domes $T_c(x)$ for Bi2212 ($T_{c12}$)\cite{Vishik} and Bi2201 ($T_{c01}$)\cite{EHud3} as green solid curves.  We note a striking correlation between the superconducting gap and the three stripe branches.  In Bi2212, ARPES finds a broad doping range where the gap is independent of doping, leading to a trisected dome model.  Notably, the ends of this flat regime fall close to the percolation limits of the $P_c=4$ stripe phase, confirming that superconductivity contributes to the special stability of the $S_0$ branch.  In contrast, for Bi2201, the superconductivity starts along the $S_0$ branch at lower doping, but then closely follows the $S_2$, or Slater branch, giving a hint why $T_c$ is so much lower.  This forms an interesting complement to the high-magnetic field studies that found 2D $P_c=$ stripes as the phase trapped on superconducting vortices.\cite{JH} 

Further data on extremely underdoped YBCO (brown curve)\cite{UDYBCO} and overdoped LSCO (blue solid curve)\cite{Boz1,Boz2} shed light on the superconducting quantum critical points and the boomerang effect.  In the underdoped regime the YBCO data extend only to $T_c=30K$, but it has been proposed that the straight line continues all the way to 90K, which would make this branch similar to the leading branch of trisected Bi2212.  We propose that it also corresponds to the $S_1$ stripe branch.  The YBCO data reveal that (1) the superfluid density $\rho_s$ reduces almost linearly to zero near $x=0.056$, and (2) in a small interval near this doping there is evidence for a superconducting quantum critical point (not illustrated in Fig.~\ref{fig:6c}).  If superconductivity only arises near the immediate interface between charge and magnetic stripes, then we can think of the $S_1$ phase as a {\it filamentary superconductor} where the superconducting filament density goes to zero at low doping, leading to both the doping and $\rho_s$ vanishing at the quantum critical point.  

Switching to the overdoped regime, we see that all three superconductors have a branch of $T_c$ starting at the end of the $S_0$ branch and falling nearly linearly to zero at a material-dependent doping near $x\sim 0.25-0.26$.  From the LSCO data, we learn that $\rho_s$ also falls linearly towards zero, and there is a narrow quantum critical region near $x=0,26$.\cite{Boz1,Boz2}  Identifying this branch with $S_2$, we again expect a filamentary eutectoid superconductor, where the filament density and $\rho_s$ both go to zero near $x=0.26$, whereas the electron density does not.  In other words, our model provides a natural explanation for both the normal Uemura plot and for the boomerang effect with overdoping.\cite{Uem1,Uem2}  Note that filamentary superconductivity has been proposed previously for cuprates\cite{FilSup}.

Let us restate the argument in more detail.  A key assumption of the eutectoid-NPS (ENPS) model is that in a clean system there are three dopings $x_A$, $x_B$, and $x_E$, where the material is single phase.  These represent the two end phases of the frustrated first-order transition, $x_A$ and $x_B$, and the doping of the eutectoid, $x_E$. For cuprates, $x_A=0$. the Mott insulator, $x_B=x_{VHS}$, associated with an LTT-like phase wave, and $x_E\sim (x_A+x_B)/2$.  If these phases are not found experimentally, the ENPS model is in trouble.  Of course, the cuprates are not clean, due to dopant clustering which leads to the patches underlying the gap maps seen in STM, but there still should be clear indications of uniform phases. 

Figure~\ref{fig:6c} shows that $x_E\sim 0.13$, and the red vertical lines indicate that it dominates the pseudogap between its percolation limits, $x_{\pm}=(x_{A/B}+x_E)/2$.  Moreover, the fact that over that whole doping range the area of the Fermi surface scarcely changes is a strong indication that this is a eutectic-like phase with fixed doping.  The clearest evidence for $x_B$ comes from the STM study finding two dominant $Q$-vectors, one for the $S_2$ density wave, but a second one at $\Gamma$.\cite{Oorbital}  This latter indicates an intra-unit-cell transition making the two oxygens inequivalent.  Seeing it experimentally means that many of the patches must share this inequivalence, with the same x-y anisotropy.  Since this is a signature of VHS nesting, it should be optimal at $x_{VHS}=x_B$.  Finally, while the AFM insulator at $x=0$ is clearly single phase, there is an extended doping range, up to $x\sim 0.03$ or 0.04 depending on the cuprate, where the Mott phase persists almost unchanged. This cutoff doping represents the Mott transition, Eq.~\ref{eq:0},\cite{MottCrit} below which the holes are bound to impurities.  The critical doping is larger than in Si or Ge, because the cuprate dielectric constant is smaller, while the hole mass is larger than in the elemental semiconductors.

Thus, there are clear experimental indications for all three fixed points.  Corresponding to the fixed points, there are three domains, $D_1:$ $x_A\le x\le x_-$, $D_0$: $x_-\le x\le x_+$, and $D_2$: $x_+\le x\le x_B$, where the $S_i$-branch dominates in domain $D_i$.  In this case, the three domains correspond to the three branches of the trisected dome.  In the $D_0$ branch the superconductivity percolates, while the other branches host a filamentary superconductivity
whose volume fraction goes to zero at the end points, $x_A$ and $x_B$.

We need just one more result to describe superconductivity, which we find by rearranging the Uemura result.
Focusing on the doping range below $x=0.125$, the Uemura relation\cite{Uem1,Uem2} can be written $T_c\sim \rho_s\sim x$, where $x$ is the doping in the stripe phases $S_0$ and $S_1$.  Now $x=x_B$(width of a charge stripe)/(distance between charge stripes)=$2x_B/P_c$.  We note two further relationships.  The stripe wave number is $Q=2\pi/P_c$, and the number of interfaces between charge and magnetic stripes, per unit length, is $N_{int}=2/P_c$.  Thus, the Uemura relation can be rewritten as
\begin{equation}
T_c \sim Q \sim N_{int}.
\label{eq:1}
\end{equation}

The advantage of this reformulation of Uemura's relation becomes clear for larger doping, on the $S_2$-branch.  This branch is the mirror of the $S_1$-branch, where the magnetic stripes act as domain walls while the charge stripes get wider.  Hence, both $Q$ and $N_{int}$ decrease with doping, thereby incorporating the boomerang effect.

Equation~\ref{eq:1} provides a strong hint to the mechanism of cuprate superconductivity.  A popular, pre-cuprate proposal for creating higher-$T_c$ superconductors\cite{Little,Ginzburg,ABBa} is to form an intimate mixture of two materials, one with strong pairing fluctuations and the other a good metal -- in the present case one Mott-like, the other Slater-like.  Notably, these ideas were used to explain an increase of $T_c$ in metal-semiconductor eutectics,\cite{Eut1,Eut2} where the semiconductors were typically Si and Ge and the metal Al.  It was found that $T_c$ can be further enhanced by quenching to make a more intimate mixture of the two end phases.  A similar Martensitic model has been proposed for superconductivity in the A15-compounds.\cite{Bilbro,VHSSC2}  
Finally, it was found that when a bilayer film is grown of La$_2$CuO$_{4+\delta}$ and heavily overdoped LSCO, an interface superconductor is produced with $T_c>  50$~K, in an interface layer about 1–2 unit cells thick.\cite{Boz_interface}

Several recent experiments suggest the correctness of this picture.  Notably, an NMR study finds strong similarities between the phase-separated superconductor in stage 4 La$_2$CuO$_{4+\delta}$, and LSCO doped at $x=0.11$, near the optimal doping of the $P_c=4$-stripe phase.\cite{Imai}  In the former sample, octahedral tilting starts below the staging onset near 290K and continues to develop down to a charge ordering onset at 60K, while spin order and superconductivity onset simultaneously at 42K.  Except for staging, the same sequence is followed in the LSCO sample.  The latter differs in that the transitions are broader and evidence for phase separation is not clearly seen.  All these features are consistent with the present model.

References~\onlinecite{MG1,MG2,MG3} demonstrated that superconducting fluctuations in most cuprates above $T_c$ cannot be explained by Ginzburg-Landau physics, but are consistent with a {\it universal} percolative superconductivity.  It was proposed\cite{MG4} that the percolation arises due to pseudogap-related inhomogeneity of the `normal' state, as in the present model.  In La$_{2-x}$Ba$_x$CuO$_4$ stripe order reduces superconductivity to a 2D form, while uniaxial stress restores 3D superconducting order\cite{smaller}.  Notably, stress does not eliminate the stripes, but only reduces their correlation length.   Finally, there is increasing evidence for the existence of pair-density waves in cuprates, in which the pair-density has stripe-like modulations.\cite{PDW1}

Equation~\ref{eq:1} provides additional support for the ENPS model.  The fact that it holds across the full pseudogap regime suggests that the interfaces for all three textures are between the same two phases, as expected for a frustrated first-order transition.  Moreover, coupling to superconductivity can account for the extended doping range where the $P_c=4$ phase is found.

We caution that current results are mainly on Bi- and La-cuprates and YBCO, for which the QOs are all from the $S_0$-branch -- i.e., on the Mott side of the Mott-Slater transition, and we have little information on the Hg- and Tl-cuprates, for which QOs indicate a larger Fermi surface, i.e., on the Slater ($S_2$) side.  For these materials even Uemura-plot studies have only been carried out on polycrystalline samples, which are considered less reliable.  For further discussion and references, see Ref.~\onlinecite{Homes}.

Indeed, many earlier studies of the overdoped regime may have been affected by sample quality limitations.  Recent technological advances have led to significantly improved samples, with charge order persisting to significantly higher doping.\cite{Tranq_beyond19,Keimer_plasmons}  This can be important to connect cuprate physics with reports of a second superconducting dome at higher dopings where a second, $d_{z^2}$-band may become involved.\cite{SCO,BCO,Johannes}

\section{Mott-Slater transition in the pseudogap phase}

\begin{figure}
\leavevmode
\rotatebox{0}{\scalebox{0.80}{\includegraphics{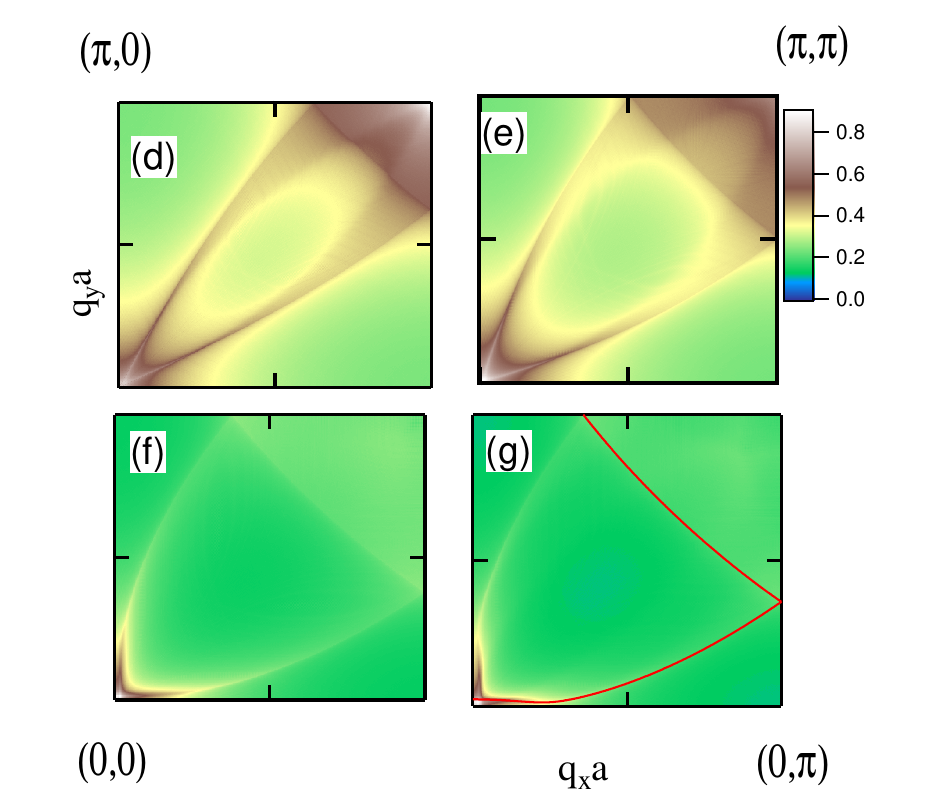}}}
\vskip0.5cm
\caption{(Color online)
Susceptibility maps at saddle-point VHS for a variety of values of $t'/t$: (d) -0.15, (e) -0.20, (f) -0.25, (g) -0.258. Red lines in (g) = folded Fermi surface nesting map, $q=2k_F$.  From Ref.~\onlinecite{hoVHS1}.
}
\label{fig:B2b}
\end{figure}

While a Mott-Slater transition in an undoped cuprate as a function of $U$ is well known, a very different, non-Landau 
 Mott-Slater transition as a function of doping or hopping parameter $t'/t$ arises in studies of the Lindhard susceptibility\cite{MBMB}.  A form of `Mottness' is a property of the bare Lindhard electronic susceptibility.  For a conventional metal in the Slater regime, the susceptibility contains a folded map of the Fermi surface, red lines in Fig.~\ref{fig:B2b}(g), most easily seen in a 2D material -- a nonanalytic peak whose intensity as a function of $q$ is a quantitative measure of the strength of nesting at that $q$-point.\cite{AIP}  However, this map is neither one-to-one nor faithful since for Fermi surface nesting $q=2k_F$, $q$ in general can extend outside the first Brillouin zone, and must be folded back.  For cuprates, this leads to a diamond-shaped ring around $(\pi,\pi)$, as seen in Fig.~\ref{fig:B2b}(f,g).  While the peak intensity is sharply defined, the intensity falls off fairly slowly away from the peak, and near $(\pi,\pi)$ the intensity from the four sides of the diamond add, and in the Mott phase lead to a diamond-shaped plateau with peak at $(\pi,\pi)$, Fig.~\ref{fig:B2b}(d,e).  At the same time, the original peaks of the map have disappeared and been replaced by a change in slope, leading to a steep falloff of intensity off of the plateau -- i.e., Fermi surface nesting is lost.  This is seen in Fig.~\ref{fig:B2b}: in frames (f) and (g) in the Slater regime, the segment of Fermi surface nesting map nearest $(\pi,\pi)$ is yellow surrounded by green -- i.e., a local maximum -- whereas in the Mott regime, frames (d) and (e), the analogous line changes from brown to yellow on crossing the line -- i.e., a step down in moving off of the Mott plateau.  As the doping or $t'$ are changed, the area of the diamond grows, reducing the overlap, and shrinking the peak at $(\pi,\pi)$, until at a critical point, the peak at $(\pi.\pi)$ turns into a local minimum, and the nesting peaks are restored -- the Mott- Slater transition.\cite{MBMB}  In Fig.~\ref{fig:B2b}, this occurs as $t'/t$ changes from -0.2 (e) to -0.25 (f).

\begin{figure}
\leavevmode
\rotatebox{0}{\scalebox{0.44}{\includegraphics{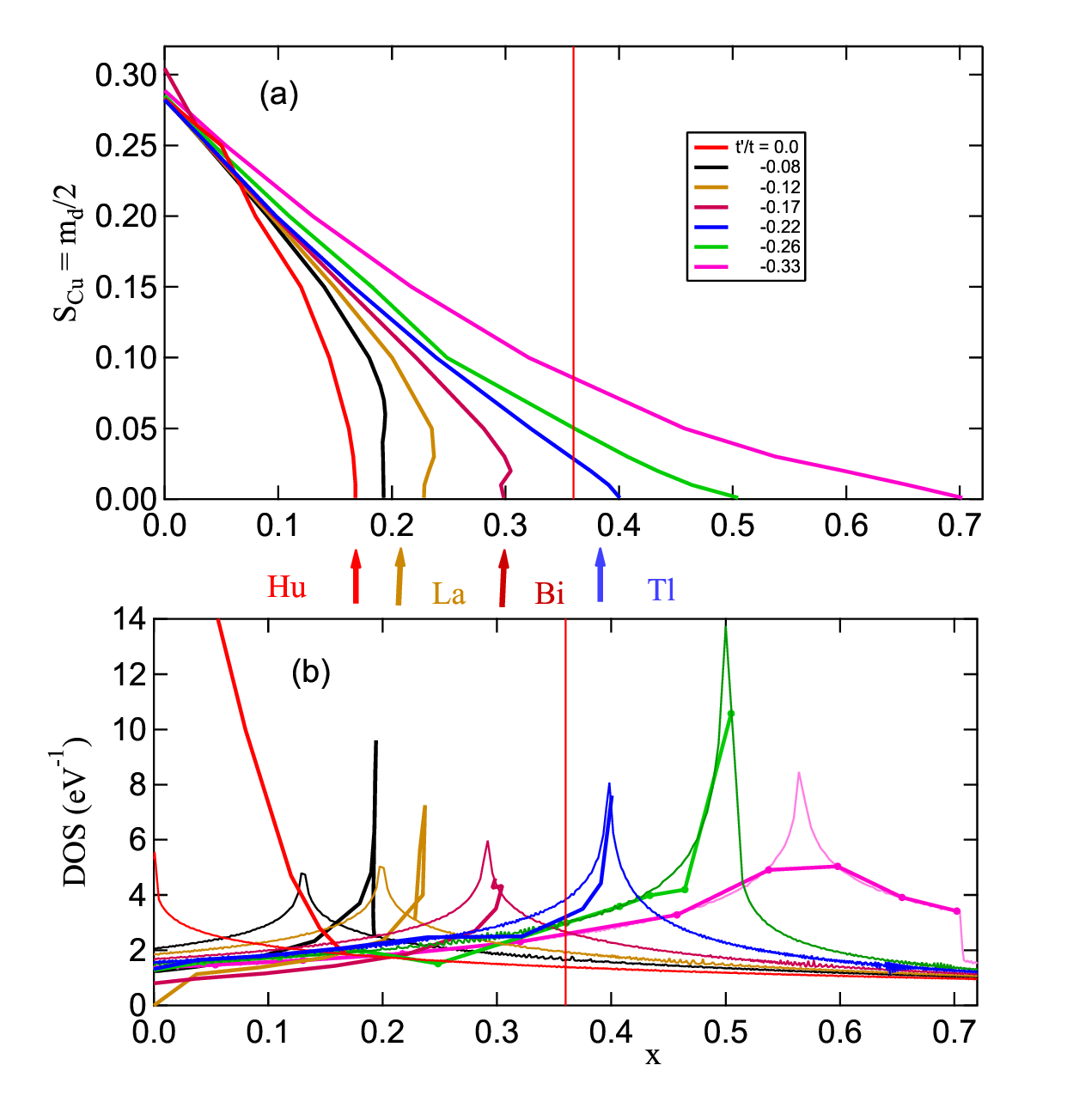}}}
\vskip0.5cm
\caption{
$(\pi,\pi)$-AFM phase in cuprates.  (a) AFM magnetization ($S$) and (b) the corresponding DOS as a function of doping ($x$), for several values of $t'$, with Hubbard $U=1.2$~eV, and renormalization factor $Z=0.5$).  In (a), the labeled arrows below the horizontal axis identify the corresponding material, with Hu for Hubbard model, La for La cuprates, Bi for Bi cuprates, and Tl for Tl cuprates.  In (b), the thin lines of the same color represent the corresponding nonmagnetic phases.  From Ref.~\onlinecite{Paper1}.
}
\label{fig:79}
\end{figure}

This transition shows up in other work as well.  Figure~\ref{fig:79} is taken from Fig.~1 of Ref.~\onlinecite{Paper1}, showing that the $t'$ dependence of the AFM magnetization, $S_{Cu}=m_d/2$, contains three distinct behaviors, one for very low $t'$ where most of the doping takes place beyond $x_{VHS}\sim 0$, an intermediate range where $S_{Cu}$ drops discontinuously to zero at a doping close to $x_{VHS}$, and a third regime where $S_{Cu}$ goes to zero continuously at a doping slightly beyond $x_{VHS}$.  Notably, the crossover between regions 2 (discontinuous) and 3 (continuous) takes place between $t'/t$ = -0.17 and -0.22, vertical red lines in Fig.~\ref{fig:79}, consistent with the Mott-Slater transition.  The reason for the discontinuity is that in region 2 the AFM DOSs are quite distinct from the NM VHSs, while in region 3 they merge smoothly onto the NM VHSs -- i.e., they are Fermi-liquid-like.  This can be seen in Fig.~\ref{fig:79}(b), where the Mott-Slater transition is close to $t'/t=-0.17$, where the AFM VHS has its smallest value.  Notably, when the transition becomes continuous the pseudogap persists beyond the VHS, $x^*\simeq x_{step}>x_{VHS}$, where $x_{step}$ indicates the doping at which the susceptibility has a discontinuous downward step\cite{Paper3}.  In the doping range between $x_{VHS}$ and $x_{step}$, there are small hole-pockets surrounding $(\pi,0)$ and $(0,\pi)$.\cite{Andersen}

We briefly note some additional signatures of the Mott-Slater transition.  The inter-VHS (at $(\pi,\pi)$) and intra-VHS (at $\Gamma$) susceptibilities are equal at $t'/t\sim -0.22$, signaling maximum competition between AFM order at $(\pi,\pi)$ and near-$\Gamma$ charge order (Fig. 4 of Ref.~\onlinecite{hoVHS1}).  We propose that this Mott-Slater transition is also reflected in the commensurate ($P_c=4$) to incommensurate ($P_c>4$) phase transition in Bi2201 near $x=0.14$,\cite{EHud2} vertical black line in SM Fig.~S1.  Thus, the $S_0$ branch is a stripe phase controlled by maximal repulsion between charge stripes, independent of the Fermi surface, while the $S_2$ branch is controlled by AFM Fermi surface nesting.  Moreover, we see in SM Fig.~S1 that the commensurate-incommensurate (C-I) transition is weakly first order, as suggested for the Mott-Slater transition.

While our analysis has been restricted to the coherent band near the Fermi level, experiments have found signatures of closely-related effects in the incoherent band at high energies, near the charge-transfer gap\cite{Mottness}.  This is related to a charge-transfer exciton that is well defined in the Mott regime, but screened out in the Slater regime, with the crossover near $x=0.16$ in Bi2201.

We note that a candidate of this doping-dependent Mott-Slater transition has also been observed in Sr$_2$Ir$_{1-x}$Rh$_x$O$_4$.\cite{SIRO4}  Finally, additional signatures are found when doping away from the VHS.  These are briefly summarized in SM IV.F.

The superconducting vortex lattice offers an analog to the Mott-Slater transition.  In most vortex lattices, the vortices form a hexagonal lattice, to keep the vortices as far apart as possible (Mott-like phase).  However, in some materials, including cuprates\cite{Vlatt1}, the vortex lattice is square or nearly so, which may be related to Fermi surface anisotropy (Slater-like phase).\cite{Vlatt2,Vlatt3}.

\section{Discussion}

\subsection{Strange metal}
While the presence of competing stripe textures and the strong scattering expected near the Mott-Slater transition make transport calculations particularly challenging, our results provide hints about the origin of the strange metal phase, defined as the part of the cuprate phase diagram dominated by a resistivity linear in $T$ and $\omega$.  

One of the earliest theoretical results for cuprates is that proximity to a VHS provides a natural explanation for such behavior\cite{PAL,RMlin}.  While realistic transport calculations find a mixture of linear-in-$T$ and $T^2$ contributions to the near-VHS resistivity\cite{Hlub,Schmitt,MFLTanmoy,YJarrell1} (see also Section~6.3.2 of Ref.~\onlinecite{VHSrev}), this is consistent with experiments, which find that the resistivity is almost never strictly linear in $T$, but some mixture of linear and quadratic, the exact form being the subject of much debate.  Moreover, in Ref.~\onlinecite{MBMB} we demonstrated that at low temperatures the effect of the VHS is limited to a small doping range due to Pauli blocking, but as $T$ increases the effects of the VHS spread over an increasing doping range as $T$ increases.  This provides a natural explanation for the characteristic geyser-like shape of the strange metal phase boundary in $x-T$ plots (SM IV.F).  Taken together, these results amount to a prediction that if the pseudogap terminates near a VHS, then phenomena similar to the strange metal phase follow automatically.  Thus, our model captures many of the experimental signatures of the strange metal phase.

However, at low temperatures the linear-in-$T$ resistivity spreads out into a finite doping range for $x>x_{VHS}$, known as the `foot'.\cite{HuHall2,Sachdev}  A possible explanation for this is discussed in SM IV.F.

\subsection{Stripes in DFT}

The new DFT calculations employing metaGGA exchange correlation functionals have shown great promise in describing AFMs\cite{metaAFMs}.  These calculations have now developed to the point that they can also describe antiphase boundaries and stripe phases in correlated materials\cite{YUBOI,Nick,Yale} and provide valuable information on their shapes and the relative importance of charge, spin, and lattice distortions.  Indeed, these studies provide a new way of looking at intertwined orders in correlated materials, one consistent with the NPS/textured framework.  Each study finds a large number of nearly degenerate phases, with the lowest energy phases having the largest magnetic moments on copper or nickel.  In YBCO$_7$, DFT finds more than 20 competing low-energy phases, mainly stripe-like, with the minimum energy phase having a charge periodicity $P_c=4a$.  Note that in a DFT calculation, each texture, or stripe configuration, counts as a separate phase, requiring a new DFT calculation with the correct unit cell.  While we only calculate a finite number of phases, these are stripe-like phases, and it is clear that as the stripe periodicity goes to infinity the phase must converge to the AFM phase.  Since the lowest energy phases have the largest magnetic moment, the driving force for stripe formation remains magnetic -- i.e., these are all textures of an underlying AFM -- even in overdoped YBCO$_7$.  Similar domain wall phases have been discussed for nickelates\cite{Nick} and ferromagnets.\cite{Coty}

As to the type II superconductor analogy, DFT calculations of the textures of the flux lattice phase in a superconductor in a magnetic field are currently not feasible.  However, Ginzburg-Landau and Bogoliubov-de Gennes calculations find very similar results with many textures arising from the extreme sensitivity to boundary conditions.

Given the existence of so many competing phases, the best way to understand the resultant physics should be to apply ideas from statistical mechanics to DFT, in particular, the idea of ensemble average.  A number of approaches have been applied.  First, one can develop a partition function, with the states being the various DFT solutions\cite{YUBOII}.  Alternatively, if one assumes ergodicity, that over time or space the material samples all its possible phases, one can do a DFT calculation on a sufficiently large unit cell, and analyze the results via molecular dynamics techniques\cite{Zung1,Zung2}.   A related self-consistent Hartree-Fock approach has been applied to nickelate superconductors\cite{WeiKu}.  These models also bear some similarity to a polymeric model of stripes.\cite{polymer}

These DFT calculations can also be used to validate our NPS stripe model.\cite{RSM2000,RSMstr}  Thus, both find similar AFM phases at half filling, and similar stripe dispersions in doped cuprates,\cite{YUBOII} Fig.~\ref{fig:11}.  Note in particular that the band dispersion develops minigaps in directions that cross the stripes, as $\Gamma\rightarrow X$ and $\Gamma\rightarrow S$, but not in the dispersion along the stripes, $\Gamma\rightarrow Y.$  Finally, in a three-band version of the model, both calculations find that AFM gaps are present both in the AB band near the Fermi level, and in the BB band nearly 6~eV away from the Fermi level.\cite{3band,SCAN2}

\subsection{Summary of Model}
We briefly outline the key features of the model.  Of all cuprates, the `exception that proves the rule' is La$_2$CuO$_{4+\delta}$.  No matter how much oxygen is added the AFM Neel temperature $T_N$ remains unchanged, while all doped samples contain a superconducting component with $T_c\sim 44$K.  This happens because the AFM is incompressible, while the intercalated oxygen is mobile near $T_N$, so all the excess oxygen forms a second phase -- a real-space example of spin-charge separation\cite{Nagaosa}.  Note that this result suggests that the eutectoid phase is host to high-$T_c$ superconductivity, consistent with our analysis of Bi2212, Section IV.  But if the sample is further doped with, e.g., Sr, which is fixed in the lattice, Coulomb effects rapidly frustrate macroscopic phase separation, and typical pseudogap behavior -- $T_N$ and $T_c$ that evolve smoothly with doping -- is quickly restored.  Hence, the goal of pseudogap theory is to understand how the cuprates adapt to this frustration while retaining a residue of spin-charge separation.

We find that doping in cuprates follows alloy theory, but mostly in the nonequilibrium NPS limit, where the doped ions are immobile.  The cuprate phase diagram is dominated by two stable end phases, an AFM insulator at half filling and a metallic nematic phase at $x_{VHS}$, with a two-phase (nanoscale) coexistence region between.  However, a mixed phase has a lower free energy than the tie-line between these two phases, leading to a eutectoid region between them.

Thus the pseudogap phase consists of an underlying AFM order with three branches involving different stripe-like phases, the eutectoid $S_0$ and two branches, one between the AFM and eutectoid ($S_1$) and one between eutectoid and nematic ($S_2$).  Primary evidence for the underlying AFM order comes from three sources: (1) DFT calculations showing that stripe phases are stabilized by optimizing the magnetic moments; (2) the finding that the $Q$-vector of the $S_2$ branch is determined by nesting of an AFM Fermi surface; and (3) the finding that several key features of pseudogap collapse are consistent with the termination of an AFM phase as the Fermi level crosses the AFM VHS.\cite{Paper1}

The resulting texture of the stripe phases can best be understood near the AFM limit: since the AFM is incompressible, doping consists of nucleating a second phase on topological defects of the AFM (i.e., antiphase boundaries).  The physics is similar to the creation of magnetic vortices in type II superconductors, and was applied earlier to CDWs\cite{KrumS}.  We note that the free energy is also anomalous near a VHS\cite{Rice,VHSrev}, with energy and entropy both diverging logarithmically, suggestive of Bereshinsky-Kosterlitz-Thouless physics.\cite{BKT}

A key result of our work is that the nematic phase is closely related to the LTT phase found in LSCO.  This suggests the possible involvement of dynamic Jahn-Teller physics in the cuprates, with connections to the thermal Hall effect and time crystals.

Finally, the eutectoid origin of the pseudogap 
suggests a connection between cuprate superconductivity and the Martensitic model of A15-superconductivity.\cite{Bilbro,VHSSC2}

\subsection{Comparison  with other many-body calculations} 
Most differences between our results and various cluster extensions of dynamical mean-field theory (DMFT) stem from two sources.  First, while most experimental groups recognize that accurate tight-binding models of cuprates require at least a finite $t''$ hopping parameter, most theories are based on Hubbard model $(t'=t''=0)$ or $t-t'$-models.  Notably, existing 3-band models contain no parameter equivalent to $t''$, although Ref.~\onlinecite{Paper3} should correct this problem.
Secondly, cluster calculations typically have coarse momentum resolution of self-energy and susceptibility, comparable to the Brillouin zone area divided by $N$, the number of clusters, typically 2-8.  In contrast, our calculations require highly accurate susceptibilities to resolve the Fermi surface map and describe the Mott-Slater transition. Notably, dynamical cluster approximation (DCA) calculations can use $N$-values as large as 56, leading to a picture of the pseudogap much closer to our model.\cite{YJarrell1}

In contrast, we find much better agreement with many-body extensions of DMFT calculations, particularly dynamic vertex approximation (D$\Gamma$A) calculations.  For instance, our results can now be compared with a recent D$\Gamma$A calculation of the 3-band Emery model of cuprates.\cite{DGA}  One expects the results to be similar, as they are based on a ``Moriyaesque'' $\lambda$-function\cite{Moresq} which satisfies Mermin-Wagner physics,\cite{MW} similar to the one we developed.\cite{RSM70,MBMB}  Indeed, they find that at low doping the pseudogap represents a short-range ordered AFM phase, with a Mott-Slater transition at finite $T$ clearly visible in the susceptibility (their Fig. 4), as predicted.\cite{MBMB} Another calculation\cite{waterfall} agrees with our model of the origin of the waterfall as due to dressing of the carriers with spin fluctuations, including the close relation to the high-energy kink, the resultant renormalization of $U$, and the need for self-consistent calculations.\cite{HEK,Susmita}

\section{Conclusions}

Correlated materials combine aspects of great complexity (intertwined orders, pseudogaps) with other aspects hinting at an underlying universality -- emergence of exotic phases with `colossal' properties (high-$T_c$ superconductivity, colossal magnetoresistance, heavy fermion behavior), and strange metals with anomalous transport (linear-in-$T$, -$\omega$ resistivity).  For cuprates, we here identify the complexity as due to the texture that confines competing orders on charge stripes, with three branches of competing stripes.  Notably, superconductivity is correlated with interfaces between magnetic and charge strpies, which provides a natural explanation for the Uemura effect, with boomerang.  We have further demonstrated that the stripe-to-CDW crossover is a defining signature of the Mott-Slater transition.  Even when the physics is controlled by Fermi surface nesting, the AFM texture can adjust to accommodate nesting, while the induced vortices ultimately melt the charge stripes, leaving the pseudogap collapse to be described by a simple AFM model.\cite{Paper1}

An advantage of the model developed here and in Ref.~\onlinecite{Paper1} is that it is readily extendable to many other correlated materials\cite{Bian3,irSCAN,niSCAN,fSCAN,Bian1,Tranq4,Bian4}, and the small correlation length limit of this model is consistent with the special quasirandom structures (SQS) model\cite{Zunger}.

\section*{Acknowledgments}     
This work was supported by the US Department of Energy (DOE), Office of Science, Basic Energy Sciences Grant No. DE-SC0022216 and benefited from the resources of Northeastern University’s Advanced Scientific Computation Center, the Discovery Cluster, the Massachusetts Technology Collaborative award number MTC-22032, the Quantum Materials and Sensing Institute, and the National Energy Research Scientific Computing Center through DOE Grant No. DE-AC02-05CH11231.

\end{document}


\renewcommand{\thefigure}{S\arabic{figure}}
\renewcommand{\figurename}{{\bf Fig.}}
\title{A model for intertwined orders in cuprates: Supplementary Material}
\author{R.S. Markiewicz, M. Matzelle, and A. Bansil}
\affiliation{Physics Department, Northeastern University, Boston MA 02115, USA}
\affiliation{Quantum Materials and Sensing Institute, Northeastern University, Burlington, MA 01803, USA} 
\maketitle

\section{The Three Textures}

\subsection{Overview and $S_1$ branch}

Here we review experiments on the pseudogap phase to ascertain a minimal picture of what a model of the cuprate pseudogap should explain.  There is evidence for two classes of competing charge order phases -- either stripe-CDW like or associated with oxygen or oxygen-vacancy order.  The former appear to be universal, and are the main topic of this section, while the latter are quite material specific and will be neglected.  
We focus mainly on the Bi-cuprates, for which the most extensive data are available.  In Fig.~1(a), we summarize the various textures that have been found in the pseudogap phase, identified by their $Q$-vectors -- specifically the charge ordering vectors found in Bi2201\cite{TranqBi,EHud2} and Bi2212\cite{per4,GhirQ}.  Certain technical details of Fig.~1(a) are discussed in SM III.A.  In SM IV, we show how these textures can explain a variety of experiments in the pseudogap.  Here we focus on classifying the various textures.

At first blush, the data from different experiments would seem to have little in common.  We find that these experiments can be interpreted in terms of three distinct textures, evolving along the lines labeled $S_0$, $S_1$ and $S_2$.  All four experiments find doping ranges involving branch $S_0$, with constant $q$-vector $Q=0.25$ (in units of $2\pi/a$), consistent with the charge periodicity $P_c=4a$ stripe phase, where $a$ is the shortest distance between copper atoms\cite{YUBOI}, but some experiments find additional charge order associated with the lines $S_1$ and $S_2$. $S_1$ is the well-known stripe $Q$ vector of the Tranquada model\cite{Tranq1}, consistent with charge stripes ordering by staying as far apart from each other as possible.  The neutron scattering data\cite{TranqBi} are consistent with the corresponding data in La-cuprates (red solid lines), where stripes are more intense. More recently, evidence for CDW order has been reported in many cuprates.  While some of this is associated with the $S_0$ branch, there is also evidence of a distinct, $S_2$ branch.

We briefly review what is known about branch $S_1$.  In the doping range below $x=0.125$ very similar neutron scattering results are found for La$_{2-x}$Sr$_x$CuO$_4$ (LSCO)\cite{Yamada}, YBa$_2$Cu$_3$O$_{6+\delta}$ (YBCO)\cite{TranqYB1,TranqYB2}, and Bi2201\cite{TranqBi}.  The doping dependence of $Q$ seems inconsistent with any possible FS nesting.  On the other hand, the $Q$ values are consistent with a purely repulsive interaction between charge stripes.  If all doped holes are assumed to go on the charged stripes, doping consists of adding more charged stripes.  Empirically, branch $S_1$ is consistent with a model of 1D stripes, with 0.5 holes associated with each domain wall.  Beyond this point, theories differ on microscopic details of the stripes and the origins of the stripe phenomenon.  We summarize the NPS model of these stripes in Section II.

\subsection{AFM Fermi surface nesting of $S_2$}

We note from Fig.~1(a), that while the $S_2$ branch was found in both scanning tunneling microscopy (STM)\cite{EHud1,EHud2} and resonant inelastic x-ray scattering (RIXS)\cite{GhirQ} experiments, and both experiments report a commensurate-incommensurate transition at the same doping $x_{C-I}\sim 0.14$ between the $S_0$ and $S_2$ branches, each experiment observes different segments  of the branches, and the data sets have only a single point in common, exactly at the transition.  We interpret this as follows. STM sees only data which are static on the time scale of the experiment, which can last several days, while RIXS sees fluctuating, quasi-elastic data.  Thus, the transition involves a ground state transition from commensurate $S_0$-stripes for $x<x_{C-I}$ to $S_2$ incommensurate stripes for $x>x_{C-I}$.  This is consistent with the results of Ref.~\onlinecite{Fate}.  This transition is enhanced by the superconductivity associated with the $S_0$ branch, Section 4.

If $S_1$ stripes and $S_2$ CDWs have independent origins, one would expect them to have very different properties -- CDWs without AFM stripes, different harmonic content, etc.  But that is not the case.  These CDWs are highly unconventional, since (1) the states in optimally doped YBCO contain both charge and spin order, and (2) the dominant contribution to minimizing their energy involves maximizing the average magnetic moment\cite{YUBOI} -- just as for the $S_1$ phase.

\subsection{$S_0$ branch: $P_c=4$ stripes and quantum oscillations}

In the cuprate pseudogap phase, the $P_c=4$ stripe phase plays a special role.  It is the commonest stripe phase found in several cuprates, responsible for signatures of a small electron-doped Fermi surface which gives rise to prominent quantum oscillations (QOs).  
Here we compare data from several different experiments to demonstrate that the doping evolution of this phase follows a percolation picture consistent with NPS. 

Figure~\ref{fig:6b} is a composite phase diagram of the many experimental transitions found in the pseudogap phase, with $T^*$ the pseudogap onset and the other transitions defined in the caption.  We focus on the Bi-cuprates, but some data from other cuprates are included.  See SM III.B for some technical details of the plotting.

\subsubsection{QOs}   

\begin{figure}
\leavevmode
\rotatebox{0}{\scalebox{1.0}{\includegraphics{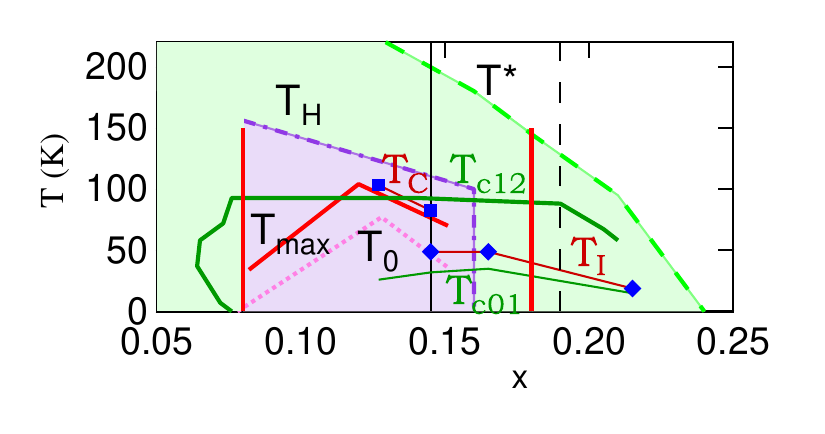}}}
\vskip0.5cm
\caption{
{\bf Pseudogap phase diagram} of YBCO\cite{Lifshitz}, compared to STM results on Bi2201\cite{EHud2}.  For YBCO, $T^*$ is the pseudogap onset, which we interpret as short-range $(\pi,\pi)$ AFM order, while we interpret $T_H$ as the Ising--X-Y transition, where domain walls become unstable, while $T_{max}$ and $T_0$ refer to the appearance of patches of charge period 4 1D (or 2D) stripes.  For Bi2201 the blue squares ($T_C$) and diamonds ($T_I$) track the peaks in the patch maps of commensurate and incommensurate stripe phases in different samples.  The green curves labelled $T_{c(n-1)n}$ are the superconducting $T_c$s of Bi22(n-1)n, $n=1,2$\cite{Vishik,EHud3}.  The red vertical lines indicate the dopings at which the QO effective mass diverges\cite{Ramshaw}, presumably associated with percolation of the $P_c=4$ stripe phase.  Solid black vertical line indicates the commensurate-incommensurate (C-I) transition, while black dashed vertical line reflects onset of interlayer coupling with increasing doping.\cite{ZXarc}.
}
\label{fig:6b}
\end{figure}

The original evidence for CDWs in cuprates came from QO measurements of a new Fermi surface, which was found to be a small electron-doped pocket, much smaller than the expected AFM pockets, with an area corresponding to 0.04 electrons per Cu.\cite{Seba}  However, this single Fermi surface is found to persist over a doping range $0.08<x<0.18$ (vertical red lines in Fig.~\ref{fig:6b}), while changing its area by only 20\%.\cite{Ramshaw}  Such behavior is not consistent with a homogeneous electronic system.  The observed area change can account for only 1/10 of the doping change, and if there is a second component that accounts for 90\% of the doping, it is highly unlikely that the small Fermi surface could avoid much larger changes.  Thus the simplest interpretation is that the $P_c=4$ phase is involved in NPS, specifically that it is uniform at precisely 1/8 doping, but away from that doping it develops topological defects (both for $x<0.125$ and $x>0.125$) that increase in number with doping.  For qualitative purposes, one can introduce a simplified percolation model with only three phases, $x_0=0$ (corresponding to doping on the magnetic stripes), $x_1=0.25$ (on the charged stripes) and $x_{QO}=(x_0+x_1)/2 =0.125$ for our stripe phase.  Then 2D percolation of the stripe phase would occur when the stripe phase population falls to 50\% of the total, or at $(x_0+x_{QO})/2=0.0625$ and $(x_1+x_{QO})/2=0.1875$, close to the experimental range\cite{Ramshaw}.  Indeed, experimental evidence for anomalies at these dopings has been proposed.\cite{Ando2}  Moreover, hopping on the percolation backbone near the limits of percolation can explain the observed effective mass divergence.\cite{Ramshaw}  In particular, to preserve a coherent QO, the electron must hop between states on the $P_c=4$ backbone, so the effective hopping parameter is $t_{eff}=tf_p$, where $f_p$ is the fraction of sites on the percolating backbone, which $\rightarrow 0$ at the percolation transition.  Thus, while for a homogeneous system, the mass renormalization factor is modest, $Z\sim 0.5$, for NPS $Z_{eff}=f_p$ can go to zero.  Such a NPS of the $P_c=4$ stripes had been predicted earlier\cite{RSMstr}.

\subsubsection{STM}
STM studies by Mesaros {\it et al.}\cite{per4} confirm this picture, finding the clear presence of patches of the $P_c=4$ phase over the full doping range.  While they ascribe this to discommensurations in a CDW phase, they don't address the issue of why this phase persists without change over such a wide doping range.  We believe that the NPS model affords a better description of the relevant physics, including percolation effects. 

Other STM groups have used masking to directly study the heterogeneity of cuprates.  It is typically found that sample surfaces consist of arrays of gap maps -- roughly 3 nm patches with distinctly different combined pseudogap-superconducting gaps covering the surface.\cite{patches}  Areas with a common gap size can be masked off and are found to correlate with a distinct  local doping\cite{EHud1}, associated with clusters of a particular interstitial oxygen impurity\cite{JennyO}.  The distinct gap sizes are associated with distinct $q$-vectors, proposed to match a nesting vector of the underlying Fermi surface\cite{EHud1} -- that is, the $q$-vector is incommensurate and changes with doping.  How does NPS reconcile these results?  Within the $P_c=4$ percolation limits, most patches have this periodicity, while the remaining patches are spread over many other periodicities.  By sorting these remaining patches, one can detect a large variety of alternate, incommensurate $q$-vectors.

In an extension of this work\cite{EHud2}, a well-defined commensurate-incommensurate transition of the stripes was found in Bi2201.  The vertical black line in Fig.~\ref{fig:6b} defines the doping $x_{CI}$, where the charge order crosses over from commensurate stripes at $P_c$=4~Cu to incommensurate $P_c>4$.  At each doping the gap size distribution was plotted, and it displays a well-defined peak.  We assume that the most common gap size correlates with the dominant domain wall order at that average doping (blue symbols in Fig.~\ref{fig:6b}  -- see SM III.B for details). The blue squares in Fig.~\ref{fig:6b} define the dopings at which a commensurate $P_c=4$ stripe order is observed, whereas the blue diamonds indicate incommensurate order, with average periodicity between $P_c=4$ and $P_c\sim 6$.  Consistent with this interpretation, we see that the blue squares persist to higher temperature at $x=1/8$, where the $P_c=4$ stripes should be a uniform phase.  We note further that at the C-I transition the gap size decreases discontinuously, so that exactly at the C-I transition the gap distribution has two peaks.  In Section VI.B we saw that this C-I transition is consistent with the Mott-Slater transition\cite{MBMB}.

The combination of an intrinsic percolative transition with extrinsic patch formation on oxygen clusters suggests a very broad distribution of $P_c=4$ patch sizes.  Clear evidence for this comes from RIXS experiments, which divided these patches by size into larger CDWs and smaller charge-density fluctuations (CDF).\cite{CDF}  These CDF are so small that it has been proposed that they give rise to an isotropic scattering that underlies the strange metal phase.\cite{Goetz_strange} 

\subsubsection{Hall effects}
Data from Hall effect experiments round out the picture of the $P_c=4$ phase.  We focus on the phase boundaries, $T_{max}$ and $T_0$, Fig.~\ref{fig:6b}, determined in a Hall effect study of YBCO\cite{Lifshitz}.  Since these boundaries are correlated with a negative component of the Hall density, they were assumed to be a signature of the same Fermi pocket found in QO studies.  Thus, $T_{max}$, which signals when the (positive) Hall coefficient $R_H$ has a maximum, is assumed to be the point at which this new phase first appears, and $T_0$, where $R_H=0$, indicates when it begins to dominate the spectrum.  The near agreement of $T_{max}$ with the STM onset temperature $T_C$, Fig.~\ref{fig:6b},  suggests the correctness of this assignment.

The high-doping end of the $P_c=4$ phase is more complicated.  First, $T_{max}$ in YBCO and $T_C$ in Bi2201 both terminate abruptly near $x=0.15$, but whereas nothing is found at higher doping in YBCO, there is an incommensurate branch $T_I$ in Bi2201.  In contrast, the QOs in YBCO extrapolate to an effective mass divergence at $x=0.18$.  One can formulate a consistent picture as follows: the $P_c=4$ phase is heading toward a percolation crossover at $x=0.18$, when this is interrupted by the C-I transition, which forms an incommensurate phase lacking the small electron pocket. Note that the last doping at which the QOs are observed is $x=0.15$, consistent with this picture. 

The remaining features in Fig.~\ref{fig:6b} are discussed in Section IV for superconductivity,\cite{Vishik,EHud3}  and in SM IV, with $T_H$, the Ising-XY transition, discussed in SM IV.B, interlayer coherence~\cite{ZXarc} in SM IV.C, and additional CDWs seen in other cuprates, possibly related to oxygen order, in SM IV.D.

\section{Many-body background}

From accurate susceptibility calculations based on detailed tight-binding model fits to density functional theory (DFT) dispersions\cite{1band}, we construct GW self energies, which renormalize the dispersion and split it into a low-energy coherent part and a high-energy incoherent part, thereby explaining the ARPES high-energy kink (HEK)\cite{AIP}.  It is found that at half filling both the coherent dispersion and the Hubbard $U$ are renormalized by a factor $Z\sim 0.5$. Notably, with this correction the coherent dispersion reproduces the renormalized ARPES dispersion.   
A mode coupling calculation\cite{MBMB} finds that $U$ is further reduced by a combination of Mermin-Wagner physics and frustration induced by competing modes.  Notably, the dominant frustration arises when the peak of the electronic susceptibility moves from commensurate $Q=(\pi,\pi)$ to incommensurate $Q=(\pi,\pi-\delta)$, passing through a very flat peak -- a Mott-Slater transition. 
The resulting moderately correlated model can be studied by self-consistent Hartree-Fock calculations.
The importance of a screened $U$ is also discussed in Ref.~\onlinecite{Unified}.

\section{Data used in constructing Figures}
\subsection{Data used to construct Fig. 1(a)}
For Fig.~1(a), we have replotted the data from Ref.~\onlinecite{GhirQ}, based on a `universal superconducting dome'\cite{UniDome}, replacing them with values based on a dome specific to Bi2201.\cite{JennySci}  Secondly, Ref.~\onlinecite{EHud2} plotted their data as $Q$ vs gap size $\Delta$ for individual gap patches.  Rather than try to convert $\Delta$ to $x$, we sketch an approximate blue line in Fig.~1(a), and only include blue circles for the average doping of each sample.  Lastly, neutron scattering\cite{TranqBi} observed magnetic nesting with $q$-vector $Q_{mx}=\pi-\delta_x$ and we inferred a corresponding charge vector $Q_x=2\delta_x$ (violet triangles).  These results are consistent with the corresponding data in La-cuprates (red solid lines), where stripes are more intense and the charge vector $Q$ has actually been observed, but only for $x$ = 0.12 and 0.15 
-- i.e., in the $P_c=4$-phase.\cite{TranqChStr}

\subsection{Data used to construct Fig. 2}

Figure~2 includes superonducting (green lines) and STM data (blue symbols), where we use the following procedure to convert the STM gaps $\Delta$ to equivalent onset temperatures.  The superconducting gaps have been well studied, and satisfy the relation 
 \begin{equation}
2\Delta=Xk_BT_c.
\label{eq:BCS} 
\end{equation}  
For Bi$_2$Sr$_2$Ca$_0$Cu$_1$O$_6$ (Bi2201), $X$ = 12.4\cite{EHud3}, while for Bi2212 $X=10$,\cite{Vishik} with weak doping dependence.  We assume all the corresponding STM gap data can be converted to temperatures by the same factor $X$.

\subsection{Data used to construct Fig.~\ref{fig:4}}
 
\begin{figure}
\rotatebox{0}{\scalebox{0.8}{\includegraphics{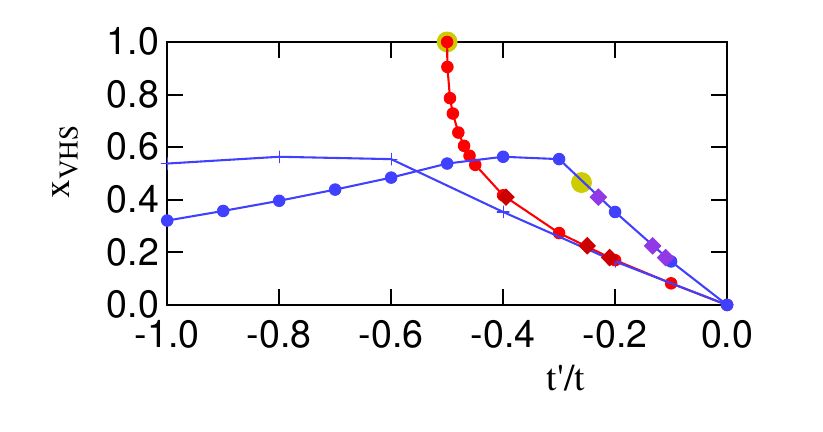}}}
\vskip0.5cm
\caption{
VHS doping $x_{VHS}$ in the $t-t'-t''$ model for $t''/t'$ = 0 (red dots) and -0.5 (blue dots).  Dark red and violet diamonds represent $t'/t$ for the data of Ref.~\onlinecite{TaillFS} within the corresponding models. Note that at low doping $t'$ for $t''=-0.5t'$ is approximately half $t'$ for $t''=0$ (blue +-signs).  Golden circles indicate hoVHSs.
}
\label{fig:A1}
\end{figure}

Ref.~\onlinecite{TaillFS} did not provide any estimate of $t'$.  To map their data onto Fig.~\ref{fig:4}, we assumed that the ARPES estimate of $x_{VHS}$ is essentially $x_{VHS0}$ and plotted the data at the appropriate $t'$.  Then Fig.~\ref{fig:A1} can be used to convert $x_{VHS0}$ to $t'$, both for the data of Refs~\onlinecite{CDMFT2,CDMFT1,Tremblay}, for which $t''=0$ (dark red diamonds), and for the parameters of the Pavarini-Andersen model, with $t''=-t'/2$ (violet diamonds). This translates into experimental\cite{TaillFS} estimates $t'_{PA}$ = -0.11 (LSCO), -0.133 (Nd-substituted LSCO), and -0.23 (Bi2201), in good agreement with other estimates.

\section{More on the pseudogap phase diagram}

\subsection{Nesting related origin of $S_2$ curve}
In Fig.~1(a) of the main text, we compare the $Q$ vector of the $S_2$ branch to a variety of models of Fermi surface nesting for Bi2201.  While the calculations for Fig.~1(a) employ a tight-binding model based on an accurate multi-parameter fit to experimental data\cite{JennySci} (the violet dot-dashed and dashed curves), we here show that similar results follow from a second model, based on a three-parameter member of the cuprate reference family\cite{MBMB} used in Ref.~\onlinecite{Paper1} (blue dashed curve) to simultaneously model all cuprates using a single tuning parameter for each cuprate family.  The NM nesting vector was taken as $Q_x=2k_{Fx}$, for the Fermi surface points $(-k_{Fx},\pi)$ and $(k_{Fx},\pi)$, leading to strong disagreement with the curve $S_2$.  We note however that the nesting $Q_x$ extrapolates to 0 at $x_{VHS}$, close to the linear extrapolation of the experimental $S_2$ curve, as predicted in the NPS model.

\begin{figure}
\rotatebox{0}{\scalebox{0.4}{\includegraphics{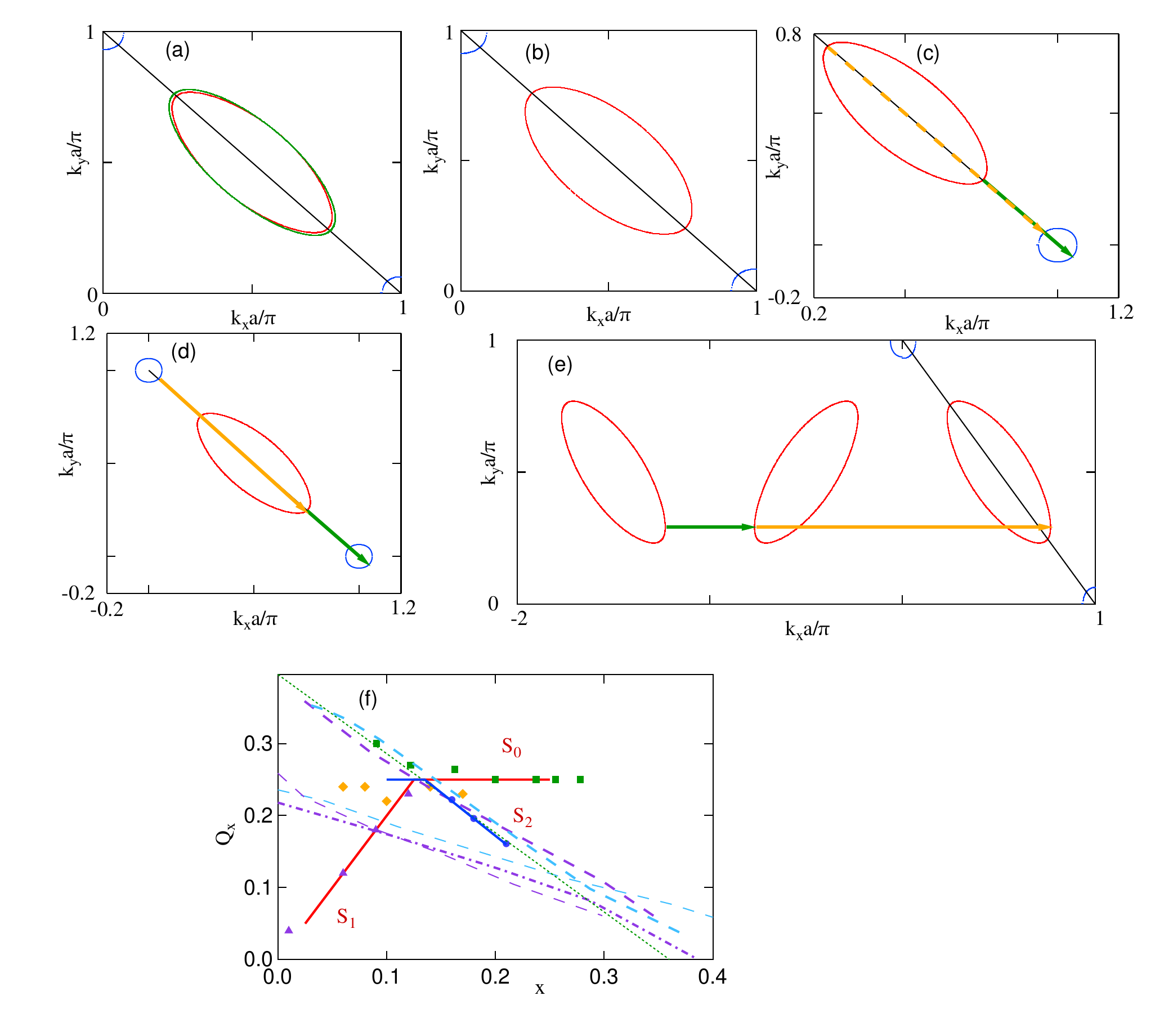}}}
\vskip0.5cm
\caption{
(a) AFM Fermi surfaces, using N-t tight binding model of Ref.~\onlinecite{JennySci} (red and blue lines) or $3-t$ model with $t'/t=-0.2$ from Ref.~\onlinecite{Paper1} (green line), with AFM gap $\Delta=150$~meV.  Fermi surfaces are electron Fermi surface from upper magnetic band (blue) and hole Fermi surface from lower magnetic band (red or green).  (b) $3-t$ model for $t'/t=-0.26$.  (c) Electron-hole nesting, with arrows denoting nesting vectors $Q'$ (green) and $S'$ (orange).  The corresponding vectors for hot-spot nesting are shown in Fig.~1(b) of the Main Text.  (d,e) Illustrating $Q+S$ for electron-hole nesting (d) and hot-spot nesting (e). (f) Replot of Fig. 1(b), with added curves: thin dashed lines: nesting curves for nesting vector of frame (c) for N-t model (violet) and 3-t model (light blue); for hot spot nesting of 3-t model (light blue thick dashed line).
}
\label{fig:A2}
\end{figure}

Recognizing the failure of models nesting the NM Fermi surface, it was proposed that nesting related to the AFM Fermi surface -- e.g., hot-spot\cite{EHud2} or Fermi arc\cite{Comin} nesting -- might prove a better starting point.  Here, we study nesting on the same AFM Fermi surface nesting model used in Ref.~\onlinecite{Paper1}, comparing the two tight-binding models noted above, Fig.~\ref{fig:A2}(a,b).  We note the resemblance to the AFM pockets in electron-doped cuprates. While there are a number of candidates for nesting, we need a $Q$-vector that starts out large at low doping and gets smaller with increased doping.  The main candidates are shown by the green lines in Figs.~\ref{fig:A2}(c) and 1(b) of the Main Text.  The electron-hole nesting in frame (c) would seem to be excluded, as the nesting vector is diagonal, whereas we need horizontal and vertical $Q$s.  However, we allow for the possibility of an exotic two-dimensional CDW, where the $x$ and $y$ nesting vectors are $Q_x$ and $Q_y$, respectively.  The two narrow dashed lines in Fig.~\ref{fig:A2}(f) represent the two tight binding models, and it is clear that neither is a good match to the data.  Instead, Fig. 1(b), the equivalent of hot-spot\cite{EHud2} or Fermi arc\cite{Comin} nesting, provides a good fit, thick violet dashed line in Fig. 1(a) and thick light blue dashed line in Fig.~\ref{fig:A2}(f).

In Figs.~\ref{fig:A2}(c) and 1(b), there is a second nesting vector, $S$ (orange arrows).   Putting both arrows on the same line, Figs.~\ref{fig:A2}(d,e), it can be seen that their sum connects a point in one Brillouin zone to the equivalent point in a different Brillouin zone, leading to the relations $S'_x+Q'_x=\pi/a$ for frame (d) and $S_x+Q_x=2\pi/a$ for frame (e).

Finally, we comment on the two tight binding models, the N-component (Fig.~\ref{fig:A2}(a) and thick violet curve in Fig.~1(a)) and the 3-component (green curve in Fig.~\ref{fig:A2}(a) and thick light-blue curve in Fig.~\ref{fig:A2}(f)).  While both models have a very similar shape of the hole Fermi surface, the electron pockets are only present for a small doping range in the three-component model, so the interpocket nesting is absent.  For a slightly larger $t'$ (Fig.~\ref{fig:A2}(b)) the electron pockets reappear, and we used this dispersion to plot interpocket nesting (thin light-blue curve in Fig.~\ref{fig:A2}(f)).  Note that both tight-binding models provide good fits to the hot-spot nesting, Fig.~\ref{fig:A2}(f).

\subsection{$T_H$ and the loss of domain walls}

Hall effect studies of YBCO\cite{Lifshitz} found a number of pseudogap-related phase boundaries, labeled $T^*$, $T_H$, $T_{max}$, and $T_0$, Figure~\ref{fig:6b}, which are believed to be representative of most cuprates (the termination of $T_H$ near $x=0.16$ is further discussed in Ref.~\onlinecite{CDWsend}). Here we focus on the first two.  If $T^*$ is the onset of short-range AFM order, as we propose, then $T_H$ has a natural interpretation as the Ising-to-XY transition\cite{Birg} at which the charge stripes melt, well below the pseudogap transition.  As the AFM phase approaches the critical doping from below, the correlation length divergence signals the broadening of the domain walls.  When the walls get broad enough, the system will transform from the Ising to an X-Y or Heisenberg model, and the walls will be destabilized.  In this range any signal of a stripe or CDW phase is lost.   This is consistent with a recent proposal that critical fluctuations of the pseudogap melt the charge order.\cite{COmelting}  The Ising-XY transition was directly observed in early experiments on cuprates\cite{Birg}.  

Similarly, the walls vanish at high doping, due to proliferation of vortices in the $S_2$ branch, so that the final pseudogap collapse can be described in terms of the primary AFM order\cite{Paper1}.

\subsection{Two dimensionality}

A recent ARPES study of Bi2212\cite{ZXarc} finds a loss of interlayer coherence below $x\sim 0.19$, specifically loss of the coupling between bilayer bands -- i.e., a transition to two-dimensionality. Notably, the transition occurs in an interesting doping regime (black dashed vertical line in Fig.~\ref{fig:6b} and black dot in Fig.~\ref{fig:4}), close to the point where the $P_c=4$ QOs first appear.
Since three-dimensionality tends to cut off the VHS DOS divergence, a loss of interlayer coherence would greatly strengthen the role of the VHS.  The loss of bilayer splitting would follow naturally if, as has been proposed, the $P_c=4$ stripes run in orthogonal directions in alternate layers.

\subsection{Other CDWs -- relation to oxygen order?}

Many additional examples of cuprate charge order are reviewed in Ref.~\onlinecite{CoDam}.  In general, the phase diagrams are similar, in that most show a dome of charge order centered near 1/8 doping, with variations in the onset temperature and width in doping of the domes.  Despite this, the $q$-vectors are very different in different cuprates.  While most are approximately consistent with the $S_1$ and $S_2$ branches, a few are notably different.  Within the present framework, this can most plausibly be understood by noting that most of these cuprates are doped by excess oxygen, and the doping sites and oxygen mobility can vary greatly in different cuprates.  Enhanced oxygen mobility can lead to larger spatial scales of phase separation, which in turn can lead to oxygen ordering with periodicity unrelated to Fermi surface nesting.

Oxygen ordering phenomena are found in a number of cuprates.  Thus, YBCO is doped by adding oxygens to a copper chain layer, and as doping varies there is a complex series of chain-ordering transitions\cite{Fontaine}.  Along these chains, STM and ARPES studies find a form of CDW order related to hole doping along the chains\cite{chainCDW}.  Similarly, in Hg cuprates evidence has been found for chain ordering of of both oxygen dopants and Hg vacancies\cite{Hgchains}.  

To study even higher doping, several groups have turned to high-pressure oxygen synthesis of LCO-analogs with the La fully replaced by Sr or Ba.  The resulting samples have turned out to have high oxygen vacancy content, Sr$_2$CuO$_{4-\delta}$ (SCO)\cite{SCO2} and Ba$_2$CuO$_{4-\delta}$ (BCO)\cite{BCO}, but with significantly increased $T_c$ values, up to 90K for SCO and 70K for BCO -- evidence for a `second dome' of superconductivity at higher doping.  Here we note only that SCO samples are multiphase, having several phases with distinct $T_c$ values, and also a number of phases with clear oxygen vacancy order.  At this point however, it is not clear whether the vacancies help or hurt superconductivity.  Oxygen vacancy ordering has also been observed in La$_{2-x}$Ca$_x$CuO$_4$.\cite{LCCO}

\subsection{Present status of superconductivity calculations}
Some models for cuprate superconductivity have reached the point of sophistication that one can begin to ask whether they can explain the observed phenomena, including the high $T_c$'s.  At this point the answers are mixed. Thus, a density matrix renormalization group (DMRG) study of the $t-t'-J$-model finds strong superconductivity for electron-doped materials but not for hole-doped -- the opposite of the cuprates.\cite{WhiSc1}  On the other hand, a DMRG study of the Hubbard model with stripes found evidence for hole superconductivity\cite{Kiv4}; note, however, that the Hubbard model is electron-hole symmetric.  A 12 cluster dynamic cluster calculation of the 2D $t-t'$ Hubbard model found clear evidence of pair field fluctuations for hole-doped cuprates, which seemed to be cut off by the pseudogap, while continuing to diverge as a Kosterlitz-Thouless transition is approached at $T_{KT}$ when the pseudogap is absent\cite{ScalMa}.  Notably, the fluctuations are absent in a $2\times 2$ cluster, which suggests that they would be hard to see in dynamic mean field calculations or their cluster extensions, which are typically limited to 8 or fewer clusters.  While Ref.~\onlinecite{Kiv1} failed to find superconductivity near a nematic quantum critical point, a followup study\cite{Kiv2} succeeded by using a stronger interaction.  In summary, it seems (1) Hubbard model results can be misleading, since the Hubbard models have an electron-hole symmetry absent in the cuprates, and breaking that symmetry can push the superconductivity towards the wrong doping.  (2) This has led to the $t-t'-J$ model being deemed unlikely for explaining cuprates, while (3) Hubbard-based models can still be viable.  Addressing the issue of whether the Hubbard model could account for a $T_c\sim 150$~K, Ref.~\onlinecite{Kiv3} concludes, ``Although affirmative answers to this question have been suggested on the basis of various approximate calculations, the presence of multiple intertwined orders—with the consequent existence of subtle energies that are difficult to capture reliably—renders the validity of these approximate results uncertain.”

Beyond calculating $T_c$, there are challenges in trying to describe the evolution of superconductivity in the presence of the pseudogap.  A particularly puzzling result is the so-called boomerang effect: that in the overdoped regime where $T_c$ is decreasing while $x$ is increasing, the superfluid density $\rho_s$ is found to also decrease and vanish near the same point that $T_c$ does.\cite{Boz1,Boz2}

However, the above calculations likely underestimate the role of the VHSs, since they are either pure Hubbard model ($t'=0$), or assume a dispersion based on a $t-t'$-model, where the VHS divergence falls at dopings far from those of most cuprates (see Ref.~\onlinecite{Paper3}).  For a cuprate-like $t-t'-t''$ dispersion, the optimal $T_c$ in various cuprate families scales with the strength of the VHS, as the VHS approaches a power law divergence\cite{hoVHS1}. However, the role of the VHS is indirect: it strengthens the Slater phase in its competition with Mott physics.  Thus, cuprate superconductivity arises not at the VHS but close to a doping of strong competition between two density wave phases, where it can control the balance of phases by piggybacking on one or the other component.\cite{hoVHS1}  This clarifies why in some materials a dome of $T_c$ is found to surround the quantum critical point of a competing order, whereas in cuprates the maximum $T_c$ generally falls in the middle of the pseudogap phase, and the superconducting dome closes near the pseudogap termination.

\subsection{More on the Mott-Slater transition}

In Section V, we showed the Mott-Slater transition as a transition in parameter space ($t'$), with crossover near $-0.22\leq t'/t\leq -0.17$.  It also arises as a function of doping, but is much less understood, with a complicated interplay of $x$ and $t'$.  Thus, doping away from the VHS, for $t'/t>-0.22$, the Mott phase persists for all dopings between half filling $x=0$ and some critical doping $x=x_M\ge x_{VHS}$, while for $t'/t<-0.22$, $x_M<x_{VHS}$ (Fig. 4 in Ref.~\onlinecite{MBMB}).  Also, for $t'/t=-0.268$ (Fig. 4 in Ref.~\onlinecite{MBMB}) and $t'/t=-0.3$ (Supplementary Fig. 3 in Ref.~\onlinecite{hoVHS1}) the system is in the Slater phase even for $x=0$.  Moreover, for $t'/t=-0.268$ the system crosses over from near-$(\pi,\pi)$ nesting to near-$\Gamma$ nesting at $x\sim 0.18$.

\begin{figure}
\leavevmode
\rotatebox{0}{\scalebox{0.70}{\includegraphics{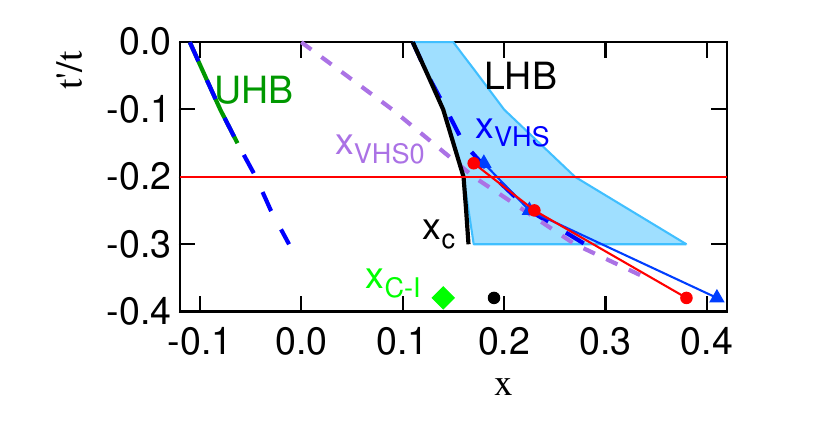}}} 
\caption{
{\bf Comparing calculated dopings of `pseudogap' $x_c$ (solid lines) and VHS $x_{VHS}$ (dashed lines), vs $t'/t$} (assuming $t''=0)$\cite{CDMFT2}.  Blue long-dashed lines indicate $x_{VHS}$ in the AFM phase, while $x_{VHS0}$ (violet short-dashed line) is the VHS of the NM phase; light-green diamond is $x_{C-I}$\cite{EHud2}; red dots and blue triangles represent experimental $x^*$ and $x_{VHS}$\cite{TaillFS}.  Black dot shows loss of interlayer coherence below $x\sim 0.19$.\cite{ZXarc}  Red horizontal line shows Mott-Slater transition, from Fig.~7.
}
\label{fig:4}
\end{figure}

When experiments\cite{TaillFS} found that the pseudogap terminates close to the doping $x$ where the VHS crosses the Fermi level, $x_{pg}\sim x_{VHS}$, several theoretical calculations\cite{CDMFT2,CDMFT1} were able to reproduce this result only for $t'>t'_{cr}$, while for $t'<t'_{cr}$, $x_{c}< x_{VHS}$, Fig.~\ref{fig:4}.  (To avoid confusion, we use the symbol $x_c$ to denote the crossover doping that was identified as the pseudogap closing in Refs.~\onlinecite{CDMFT2,CDMFT1}.)

To compare these results with experiment\cite{TaillFS}, we plot Fig.~2 of Ref.~\onlinecite{CDMFT2} in Fig.~\ref{fig:4}, making use of a symmetry of the Hubbard model (under $x\rightarrow -x$ and $t'\rightarrow -t'$) to replot the data for $t'>0$, $x>0$ (hole doped, non-cuprate like) as $t'<0$, $x<0$ (electron doped cuprate-like). We first focus on the comparison between $x_{c}$ (solid black and green lines) and the dressed $x_{VHS}$ (long-dashed blue lines).  These are in good agreement for electron-doping ($x<0)$ into the upper Hubbard band (UHB), but for hole-doping into the LHB the agreement lasts only for $t'>t'_c\sim$~-0.15, while for smaller $t'$, $x_c<x_{VHS}$.  However, $x_c$ is in poor agreement with the experimental\cite{TaillFS} $x^*$ values (red circles), which are much closer to $x^*=x_{VHS}$, as discussed in Ref.~\onlinecite{Paper1}.  (We discuss in SM-II.D how the experimental data are plotted here.)  

What is the cause for this discrepancy?
We suggest one possibility.  The deviation $x_C<x_{VHS}$ starts when $t'/t$ crosses the Mott-Slater transition, red line in Fig.~\ref{fig:4}.  This is known to cause a strong reduction in the correlation length, which should show up in experiments or theoretical calculations.   Thus, $x_c$ is in reasonable agreement with $x_{C-I}$ for Bi2201 (green diamond in Fig.~\ref{fig:4}).  Moreover, many early experiments suggested that the pseudogap terminated at $x^*=0.19$ universally in cuprates\cite{TallStorey}, whereas more recent experiments find a strong doping dependence for $x^*$ (red dots in Fig.~\ref{fig:4}).  Thus, the early experiments may have been reporting $x_C$.

The region between $x_{C-I}$ and $x_{VHS}$ is likely to be highly disordered, both because of frustration due to competing phases and reduced correlation length in the incommensurate regime, because of the emergent spin liquid phase near the C-I transition, and because of strong scattering associated with the VHS.  Hence the results of Ref.~\onlinecite{Tremblay} offer some confirmation of the current assignments.  There it is found that the light-blue shaded region in Fig.~\ref{fig:4} is a strange metal Fermi liquid phase, with linear-in-$T$ resistivity.  This is consistent with our observation, Section V.B, that the strange metal phase lies close to the VHS, where a resistivity linear in $T$ and $\omega$ has long been expected.\cite{PAL,RMlin}  

We note the development of a `foot' in the strange metal phase starting near $x_{C-I}$, rather than displaying linear-in-$T$ resistivity only near $x_{VHS}$. We suggest that this may be related to the recovery of interlayer coupling in this regime (black dot), which will cut off the VHS divergence, leaving a large but finite DOS peak spread over a finite doping range.